\documentclass[10pt, conference]{IEEEtran}
\IEEEoverridecommandlockouts
\usepackage[noadjust]{cite}
\usepackage{amsmath,amssymb,amsfonts}
\usepackage[ruled,vlined]{algorithm2e}
\usepackage{algorithmic}
\usepackage{graphicx}
\usepackage{textcomp}
\usepackage{todonotes}
\usepackage{xcolor}
\allowdisplaybreaks

\usepackage[nowarn,acronyms,nonumberlist,nopostdot,nomain,nogroupskip]{glossaries}
\usepackage{xcolor}
\newacronym{3gpp}{3GPP}{3rd Generation Partnership Project}
\newacronym{5g}{5G}{5\textsuperscript{th} Generation}
\newacronym{5gc}{5GC}{5G Core}
\newacronym{bs}{BS}{Base Station}
\newacronym{abft}{A-BFT}{Association-BeamForming Training}
\newacronym[firstplural=Access Categories (ACs)]{ac}{AC}{Access Category}
\newacronym{adc}{ADC}{Analog to Digital Converter}
\newacronym{addts}{ADDTS}{Add Traffic Stream}
\newacronym{afbw}{AFBW}{Average Fading Bandwidth}
\newacronym{aid}{AID}{Association Identifier}
\newacronym{aimd}{AIMD}{Additive Increase Multiplicative Decrease}
\newacronym{am}{AM}{Acknowledged Mode}
\newacronym{amc}{AMC}{Adaptive Modulation and Coding}
\newacronym{ampdu}{A-MPDU}{MAC Protocol Data Unit Aggregation}
\newacronym{aoa}{AoA}{Angle of Arrival}
\newacronym{aod}{AoD}{Angle of Departure}
\newacronym{ap}{AP}{Access Point}
\newacronym{app}{APP}{Application Layer}
\newacronym{aqm}{AQM}{Active Queue Management}
\newacronym{ar}{AR}{Augmented Reality}
\newacronym{ati}{ATI}{Announcement Transmission Interval}
\newacronym{awgn}{AGWN}{Additive White Gaussian Noise}
\newacronym{awv}{AWV}{Antenna Weight Vector}
\newacronym{balia}{BALIA}{Balanced Link Adaptation}
\newacronym{bdp}{BDP}{Bandwidth-Delay Product}
\newacronym{bf}{BF}{Beamforming}
\newacronym{bhi}{BHI}{Beacon Header Interval}
\newacronym{bi}{BI}{Beacon Interval}
\newacronym{brp}{BRP}{Beam Refinement Protocol}
\newacronym{bss}{BSS}{Basic Service Set}
\newacronym{bti}{BTI}{Beacon Transmission Interval}
\newacronym{cad}{CAD}{Computer-aided Design}
\newacronym{cbap}{CBAP}{Contention-Based Access Period}
\newacronym{cbr}{CBR}{Constant Bitrate}
\newacronym{cc}{CC}{Congestion Control}
\newacronym{cdf}{CDF}{Cumulative Distribution Function}
\newacronym{cir}{CIR}{Channel Impulse Response}
\newacronym{cn}{CN}{Core Network}
\newacronym{cnn}{CNN}{Convolutional Neural Network}
\newacronym{cp}{CP}{Control Plane}
\newacronym{cqi}{CQI}{Channel Quality Indicator}
\newacronym{crs}{CRS}{Cell Reference Signal}
\newacronym{csirs}{CSI-RS}{Channel State Information - Reference Signal}
\newacronym{csmaca}{CSMA/CA}{Carrier Sense Multiple Access with Collision Avoidance}
\newacronym{cts}{CTS}{Clear to Send}

\newacronym{dc}{DC}{Dual Connectivity}
\newacronym{dce}{DCE}{Direct Code Execution}
\newacronym{dcf}{DCF}{Distributed Coordination Function}
\newacronym{dci}{DCI}{Downlink Control Information}
\newacronym{delts}{DELTS}{Delete Traffic Stream}
\newacronym{dked}{DKED}{Double Knife Edge Diffraction}
\newacronym{dl}{DL}{Downlink}
\newacronym{dmg}{DMG}{Directional Multi-Gigabit}
\newacronym{dmr}{DMR}{Deadline Miss Ratio}
\newacronym{dmrs}{DMRS}{DeModulation Reference Signal}
\newacronym{dnn}{DNN}{Deep Neural Network}
\newacronym{dqn}{DQN}{Deep Q-Network}
\newacronym{dti}{DTI}{Data Transmission Interval}
\newacronym{e2e}{E2E}{End-to-End}
\newacronym{ecn}{ECN}{Explicit Congestion Notification}
\newacronym{edca}{EDCA}{Enhanced Distributed Channel Access}
\newacronym{edf}{EDF}{Earliest Deadline First}
\newacronym{enb}{eNB}{evolved Node Base}
\newacronym{endc}{EN-DC}{E-UTRAN-\gls{nr} \gls{dc}}
\newacronym{epc}{EPC}{Evolved Packet Core}
\newacronym{es}{ES}{Edge Server}
\newacronym{ese}{ESE}{Extended Schedule Element}
\newacronym{fdd}{FDD}{Frequency Division Duplexing}
\newacronym{fdma}{FDMA}{Frequency Division Multiple Access}
\newacronym{fov}{FoV}{Field-of-View}
\newacronym{fs}{FS}{Fast Switching}
\newacronym{ftp}{FTP}{File Transfer Protocol}
\newacronym{gnb}{gNB}{Next Generation Node Base}
\newacronym{harq}{HARQ}{Hybrid Automatic Repeat reQuest}
\newacronym{hetnet}{HetNet}{Heterogeneous Network}
\newacronym{hh}{HH}{Hard Handover}
\newacronym{hol}{HOL}{Head-of-Line}
\newacronym{hqf}{HQF}{Highest-quality-first}
\newacronym{ia}{IA}{Initial Access}
\newacronym{iab}{IAB}{Integrated Access and Backhaul}
\newacronym{ibss}{IBSS}{Independent Basic Service Set}
\newacronym{id}{ID}{Identifier}
\newacronym{imt}{IMT}{International Mobile Telecommunication}
\newacronym{inr}{INR}{Interference to Noise Ratio}
\newacronym{iot}{IoT}{Internet of Things}
\newacronym{ipa}{IPA}{Inter-Packet Arrival}
\newacronym{ism}{ISM}{Industrial, Scientific, and Medical}
\newacronym{kpi}{KPI}{Key Performance Indicator}
\newacronym{lcf}{LCF}{Level Crossing Frequency}
\newacronym{lcr}{LCR}{Level Crossing Rate}
\newacronym{los}{LoS}{Line-of-Sight}
\newacronym{lp}{LP}{Low Power}
\newacronym{lte}{LTE}{Long Term Evolution}
\newacronym{m2m}{M2M}{Machine to Machine}
\newacronym{mac}{MAC}{Medium Access Control}
\newacronym{mc}{MC}{Multi-Connectivity}
\newacronym{mcs}{MCS}{Modulation and Coding Scheme}
\newacronym{mdp}{MDP}{Markov Decision Process}
\newacronym{mec}{MEC}{Mobile Edge Cloud}
\newacronym{mi}{MI}{Mutual Information}
\newacronym{mib}{MIB}{Master Information Block}
\newacronym{mimo}{MIMO}{Multiple Input, Multiple Output}
\newacronym{mumimo}{MU-MIMO}{Multi-User Multiple Input, Multiple Output}
\newacronym{ml}{ML}{Machine Learning}
\newacronym{mlr}{MLR}{Maximum-local-rate}
\newacronym[plural=\gls{mme}s,firstplural=Mobility Management Entities (MMEs)]{mme}{MME}{Mobility Management Entity}
\newacronym{mmw}{mmW}{Millimeter Wave}
\newacronym{mmwave}{mmWave}{Milimeter Wave}
\newacronym{moi}{MoI}{Method of Images}
\newacronym{mpc}{MPC}{Multi Path Component}
\newacronym{mptcp}{MPTCP}{Multipath TCP}
\newacronym{mr}{MR}{Maximum Rate}
\newacronym{mrdc}{MR-DC}{Multi \gls{rat} \gls{dc}}
\newacronym{mrt}{MRT}{Maximum Ratio Transmission}
\newacronym{mss}{MSS}{Maximum Segment Size}
\newacronym{mtd}{MTD}{Machine-Type Device}
\newacronym{mtu}{MTU}{Maximum Transmission Unit}
\newacronym{nav}{NAV}{Network Allocation Vector}
\newacronym{ncbr}{NCBR}{Non-Constant Bitrate}
\newacronym{nfv}{NFV}{Network Function Virtualization}
\newacronym{nlos}{NLoS}{Non-Line-of-Sight}
\newacronym{nr}{NR}{New Radio}
\newacronym{nrmse}{NRMSE}{Normalized Root Mean Square Error}
\newacronym{ns3}{ns-3}{Network Simulator 3}
\newacronym{nsa}{NSA}{Non Stand Alone}
\newacronym{o2i}{O2I}{Outdoor-to-Indoor}
\newacronym{ofdm}{OFDM}{Orthogonal Frequency Division Multiplexing}
\newacronym{pa}{PA}{Position-aware}
\newacronym{pan}{PAN}{Personal Area Network}
\newacronym{pbch}{PBCH}{Physical Broadcast Channel}
\newacronym{pbss}{PBSS}{Personal Basic Service Set}
\newacronym{pci}{PCI}{Physical Cell Identity}
\newacronym{pcp}{PCP}{\gls{pbss} Central Point}
\newacronym{pcpap}{PCP/AP}{\acrlong{pcp}/\acrlong{ap}}
\newacronym{pdcch}{PDCCH}{Physical Downlonk Control Channel}
\newacronym{pdcp}{PDCP}{Packet Data Convergence Protocol}
\newacronym{pdsch}{PDSCH}{Physical Downlink Shared Channel}
\newacronym{pdu}{PDU}{Packet Data Unit}
\newacronym{pdf}{PDF}{Probability Distribution Function}
\newacronym{pf}{PF}{Proportional Fair}
\newacronym{pgw}{PGW}{Packet Gateway}
\newacronym{phy}{PHY}{Physical Layer}
\newacronym{ppp}{PPP}{Poisson Point Process}
\newacronym{prb}{PRB}{Physical Resource Block}
\newacronym{pss}{PSS}{Primary Synchronization Signal}
\newacronym{pucch}{PUCCH}{Physical Uplink Control Channel}
\newacronym{pusch}{PUSCH}{Physical Uplink Shared Channel}
\newacronym{qd}{QD}{Quasi Deterministic}
\newacronym{qos}{QoS}{Quality of Service}
\newacronym{rach}{RACH}{Random Access Channel}
\newacronym{ran}{RAN}{Radio Access Network}
\newacronym[firstplural=Radio Access Technologies (RATs)]{rat}{RAT}{Radio Access Technology}
\newacronym{red}{RED}{Random Early Detection}
\newacronym{rf}{RF}{Radio Frequency}
\newacronym{ris}{RIS}{Reconfigurable Intelligent Surfaces}
\newacronym{rl}{RL}{Reinforcement Learning}
\newacronym{rlc}{RLC}{Radio Link Control}
\newacronym{rlf}{RLF}{Radio Link Failure}
\newacronym{rnn}{RNN}{Recurrent Neural Network}
\newacronym{rr}{RR}{Round Robin}
\newacronym{rrc}{RRC}{Radio Resource Control}
\newacronym{rrm}{RRM}{Radio Resource Management}
\newacronym{rs}{RS}{Remote Server}
\newacronym{rsrp}{RSRP}{Reference Signal Received Power}
\newacronym{rsrq}{RSRQ}{Reference Signal Received Quality}
\newacronym{rss}{RSS}{Received Signal Strength}
\newacronym{rssi}{RSSI}{Received Signal Strength Indicator}
\newacronym{rt}{RT}{Ray Tracer}
\newacronym{rts}{RTS}{Request to Send}
\newacronym{rtt}{RTT}{Round Trip Time}
\newacronym{rw}{RW}{Receive Window}
\newacronym{rx}{RX}{Receiver}
\newacronym{sa}{SA}{standalone}
\newacronym{sack}{SACK}{Selective Acknowledgment}
\newacronym{sap}{SAP}{Service Access Point}
\newacronym{sc}{SC}{Single Carrier}
\newacronym{sca}{SCA}{Successive Convex Approximation}
\newacronym{sch}{SCH}{Secondary Cell Handover}
\newacronym{scm}{SCM}{Spatial Channel Model}
\newacronym{scoot}{SCOOT}{Split Cycle Offset Optimization Technique}
\newacronym{sdma}{SDMA}{Spatial Division Multiple Access}
\newacronym{sdr}{SDR}{Software Defined Radio}
\newacronym{si}{SI}{Study Item}
\newacronym{sib}{SIB}{Secondary Information Block}
\newacronym{sinr}{SINR}{Signal-to-Interference-plus-Noise Ratio}
\newacronym{sir}{SIR}{Signal-to-Interference Ratio}
\newacronym{sls}{SLS}{Sector-Level Sweep}
\newacronym{sm}{SM}{Saturation Mode}
\newacronym{snr}{SNR}{Signal-to-Noise Ratio}
\newacronym{son}{SON}{Self-Organizing Network}
\newacronym{sp}{SP}{Service Period}
\newacronym{spr}{SPR}{Service Period Request}
\newacronym{srs}{SRS}{Sounding Reference Signal}
\newacronym{ss}{SS}{Synchronization Signal}
\newacronym{ssb}{SSB}{\gls{ss}}
\newacronym{sss}{SSS}{Secondary Synchronization Signal}
\newacronym{ssw}{SSW}{Sector Sweep}
\newacronym{sta}{STA}{Station}
\newacronym{stb}{STB}{Set Top Box}
\newacronym{tb}{TB}{Transport Block}
\newacronym{tcp}{TCP}{Transmission Control Protocol}
\newacronym{tdd}{TDD}{Time Division Duplexing}
\newacronym{tdma}{TDMA}{Time Division Multiple Access}
\newacronym{tfl}{TfL}{Transport for London}
\newacronym{tgad}{TGad}{Task Group ad}
\newacronym{tgay}{TGay}{Task Group ay}
\newacronym{tsconst}{TSCONST}{Traffic Scheduling Constraint}
\newacronym{tm}{TM}{Transparent Mode}
\newacronym{trp}{TRP}{Transmitter Receiver Pair}
\newacronym{ts}{TS}{Traffic Stream}
\newacronym{tspec}{TSPEC}{Traffic Specification}
\newacronym{tti}{TTI}{Transmission Time Interval}
\newacronym{ttt}{TTT}{Time-to-Trigger}
\newacronym{tx}{TX}{Transmitter}
\newacronym[firstplural=Transmission Opportunities (TXOPs)]{txop}{TXOP}{Transmission Opportunity}
\newacronym{uav}{UAV}{Unmanned Aerial Vehicle}
\newacronym{udp}{UDP}{User Datagram Protocol}
\newacronym{ue}{UE}{User Equipment}
\newacronym{ul}{UL}{Uplink}
\newacronym{um}{UM}{Unacknowledged Mode}
\newacronym{uma}{UMa}{Urban Macro}
\newacronym{uml}{UML}{Unified Modeling Language}
\newacronym{upa}{UPA}{Uniform Planar Square Array}
\newacronym{utc}{UTC}{Urban Traffic Control}
\newacronym{vbr}{VBR}{Variable Bit Rate}
\newacronym{vm}{VM}{Virtual Machine}
\newacronym{vr}{VR}{Virtual Reality}
\newacronym{wbf}{WBF}{Wired Bias Function}
\newacronym{wf}{WF}{Wired-first}
\newacronym{wifi}{Wi-Fi}{Wireless Fidelity}
\newacronym{wigig}{WiGig}{Wireless Gigabit}
\newacronym{wlan}{WLAN}{Wireless Local Area Network}
\newacronym{ber}{BER}{Bit Error Rate}
\newacronym{arf}{ARF}{Auto Rate Fallback}
\newacronym{semm}{SEMM}{SPCA-EDCA Mixed Mode}
\newacronym{ppdu}{PPDU}{PHY Protocol Data Unit}
\newacronym{xlmimo}{XL-MIMO}{extra large scale massive MIMO}
\def\BibTeX{{\rm B\kern-.05em{\sc i\kern-.025em b}\kern-.08em
    T\kern-.1667em\lower.7ex\hbox{E}\kern-.125emX}}

\begin{document}

\title{Energy-Efficient Design for RIS-assisted UAV communications  in beyond-5G Networks
\thanks{This work has been submitted to IEEE for possible publication. Copyright may be transferred without notice, after which this version may no longer be accessible.
This project has received funding from the European Union’s Horizon 2020 research and innovation programme under the Marie Skłodowska-Curie Grant agreement No. 813999.}
}

\author{Anay Ajit Deshpande$^{1,*}$, Cristian J. Vaca-Rubio$^2$, Salman Mohebi$^1$, Dariush Salami$^3$,\\ Elisabeth de Carvalho$^2$, Petar Popovski$^2$, Stephan Sigg $^3$, Michele Zorzi$^1$, Andrea Zanella$^1$ \\
$^1$Department of Information Engineering, University of Padova, Padova, Italy \\
$^2$Department of Electronic Systems, Aalborg University, Aalborg, Denmark \\
$^3$Department of Communications and Networking, Aalto University, Espoo, Finland\\
\{deshpande$^*$,mohebi,zorzi,zanella\}@dei.unipd.it, \{cjvr,edc,petarp\}@es.aau.dk, \{dariush.salami, stephan.sigg\}@aalto.fi\\
$^*$Corresponding Author
}

\maketitle

\begin{abstract}
The usage of Reconfigurable Intelligent Surfaces (RIS) in conjunction with Unmanned Ariel Vehicles (UAVs) is being investigated as a way to provide energy-efficient communication to ground users in dense urban areas. In this paper, we devise an optimization scenario to reduce overall energy consumption in the network while guaranteeing certain Quality of Service  (QoS) to the ground users in the area.
Due to the complex nature of the optimization problem, we provide a joint UAV trajectory and RIS phase decision to minimize transmission power of the UAV and Base Station (BS) that yields good performance with lower complexity. So, the proposed method uses a Successive Convex Approximation (SCA) to iteratively determine a joint optimal solution for UAV Trajectory, RIS phase and BS and UAV Transmission Power. The approach has, therefore, been analytically evaluated under different sets of criterion.

\end{abstract}


\section{Introduction} \label{sec:introduction}
Increasing demand for sustainable and flexible connectivity specifically for either semi-urban/rural areas~\cite{munaye2020resource,liu2021reconfigurable} or disaster scenarios for monitoring and surveillance~\cite{olsson2010generating,luo2019unmanned}, has led to focus on the usage of \glspl{uav} and \gls{ris} for enhancing the network coverage and, thereby, the service availability of cellular networks. The conceptual design of \gls{ris} consists of several reflective elements which can be configured so as to reflect and, in particular, beamform a signal towards a particular direction. The idea of incorporating \glspl{uav} and \gls{ris} has gained traction in the last couple of years. Recently, there have been certain works that have provided definitions and optimization scenarios to tackle the direct links between \glspl{uav} and \glspl{ue} as well as links between \gls{uav} and \gls{ue} with the aid of \gls{ris}~\cite{cai2020resource, li2020reconfigurable, ranjha2020urllc, yang2020performance}.
%explain the techniques and scenarios in the papers
However, the issues of the existence and capacity limitation of the link from \glspl{bs} to \glspl{uav} have not been considered so far in conjunction with the issue of optimizing the \gls{uav} movement and \gls{ris} configuration. This should not be overlooked as the performance of the system clearly depends on the whole path from \gls{bs} to the \glspl{ue}.  Using \glspl{uav} and \gls{ris} in conjunction increases the network flexibility and makes it possible to dynamically reconfigure the system based on the network load and service requirements. Indeed, \glspl{uav} and \gls{ris} can be used to create mobile micro cells to serve temporary hotspots, i.e., areas with very high service requirement at a certain time. Additionally, the joint usage of \glspl{uav} and \gls{ris} can also enable to learn and adapt the network based on information such as user mobility and density to satisfy the user service requirements~\cite{sheen2021deep, li2019board, yi2020deep}. Additionally, the availability of high frequency communication technologies, such as mmWave~\cite{colpaert20203d}, to satisfy the higher bandwidth requirements in beyond 5G networks has increased the interest on exploiting the unconstrained mobility of \glspl{uav} to provide dynamic coverage where and when needed. Also, the enhanced beamforming capabilities of \gls{ris} can be exploited to increase the coverage for mmWave networks~\cite{nemati2020ris}. These new technologies have individually provided significant improvement in terms of service availability in semi-urban/rural areas or disaster scenarios, while potentially reducing the energy consumption of the system~\cite{fotouhi2019survey}. But the usage of high frequency technologies in conjunction with both \gls{uav} and \gls{ris} raises new challenges in terms of network optimization. In particular, a significant issue regards the trade-off between communication range and quality of service in a dense urban scenario.\\

One of the major hurdles while using both these technologies is the energy consumption of the system as a whole. \glspl{uav}, especially quadcopters, generally run on small batteries and the energy consumption is very high when the \gls{uav} is in flight. Therefore, to provide sustained coverage to the \glspl{ue} with high \gls{qos} requirements, the trajectory of the \gls{uav} has to be optimized. The use of \gls{ris}, which can improve the coverage in certain areas, may help reducing the need for \glspl{uav} to travel further, with a small trade-off on the energy consumed for \gls{ris} operation~\cite{cai2020resource, huang2019reconfigurable, abeywickrama2018empirical}. Additionally, to the best of the author's knowledge, due to the absence of power consumption model for \gls{ris}, the power consumed is supposed to be constant over a period of time~\cite{huang2019reconfigurable}. Therefore, the parametrization of the energy consumed by \gls{ris} reconfiguration and its inclusion into the energy minimization problem is thereby left for future work.\\
%As the energy consumption in \gls{ris} is still a research problem, we need to parametrize it by assuming $E^{RIS}$ joules per unit for reconfiguration, with the goal of finding how small the energy should be in order to obtain an overall net benefit.\\

\begin{figure}
    \centering
    \includegraphics[width=0.45\textwidth]{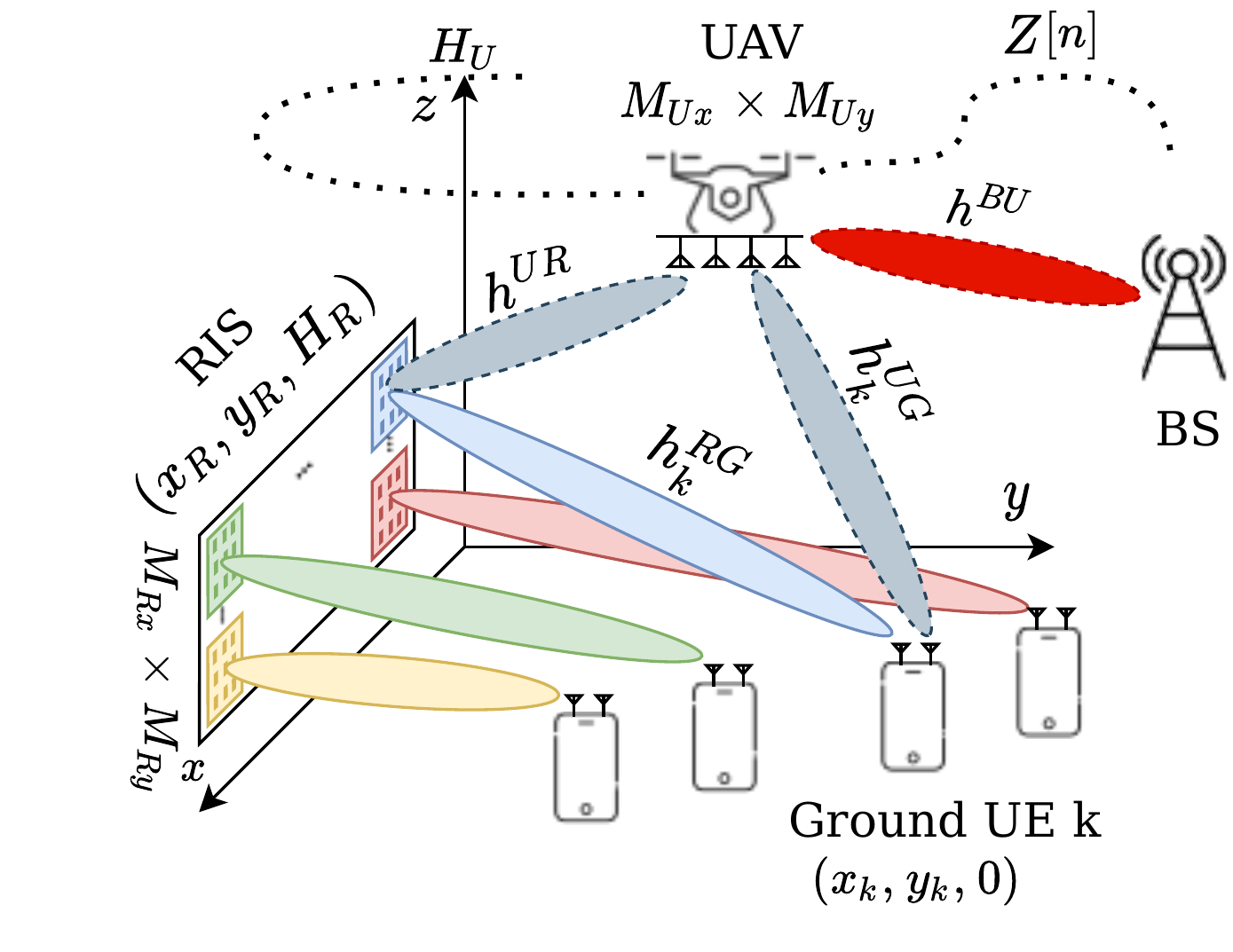}
    \caption{The problem scenario}
    \label{fig:problem_scanrio}
\end{figure}
To summarize, in this paper we explore the possibility of the combined usage of \glspl{uav} and \glspl{ris} to reduce the energy consumption of the entire system, while providing a certain level of \gls{qos} to the \glspl{ue} in the area. \figurename~\ref{fig:problem_scanrio} denotes the overall scenario in question. The \gls{uav} acts as a mobile \gls{bs} relay that can establish \gls{los} links with the \glspl{ue} and the \gls{ris}, something that might not be always possible for the fixed \gls{bs}.The \gls{uav} hence extends the area of coverage of the \gls{bs}, while optimizing the energy consumption for in-flight movement and signal transmission due to the inclusion of the \gls{ris}. If the \gls{ris} position is optimal, which is another open research problem, the \glspl{ue} can be served either directly by the \gls{uav} or with the help of the \gls{ris} or therefore combination of both, potentially reducing the energy consumption for in-flight movement of the \glspl{uav}. This can potentially extend the area of coverage (i.e., of the area of satisfactory \gls{qos}), also in situations where a \gls{bs} could not be relied upon for service, such as emergency or disaster scenarios~\cite{luo2019unmanned}. 

The contributions of this work are:
\begin{itemize}
    \item Defining a scenario and solving the associated optimization problem with respect to \gls{uav} trajectory, \gls{ris} phase shift and \gls{bs}-\gls{uav} link capacity limitation due to the \gls{uav} motion to provide at least a minimum guaranteed \gls{qos} to the \glspl{ue}.
    \item Minimization of the transmission power of \gls{uav} and \gls{bs} by jointly optimizing the \gls{uav} trajectory and the \gls{ris} phase shift.
\end{itemize}
The paper is structured as follows: Sec.~\ref{sec:introduction} provides introduction and motivation for the usage of \glspl{uav} and \gls{ris} in cellular networks. Sec.~\ref{sec:problem formulation} explains the optimization scenario and provides a brief formulation of the optimization problem whose solution is outlined in Sec.~\ref{sec:solution}. Sec.~\ref{sec:results} reports the simulation results for the obtained solution and the related discussion. Sec.~\ref{sec:conclusion} provides the conclusion and future research directions.

\paragraph*{Notations} Italic lowercase letter \textit{a} is a scalar. $\Vert a\Vert$ is a norm-two of a vector. $(\cdot)^T$ and $(\cdot)^H$ are transpose and Hermitian (conjugate transport), respectively. $\otimes$ is a Kronecker product. $\mathbb{C}$ is the complex numbers set.

\section{Scenario Definition and Problem Formulation}\label{sec:problem formulation}

\begin{figure}
    \centering
    \includegraphics[width=0.45\textwidth]{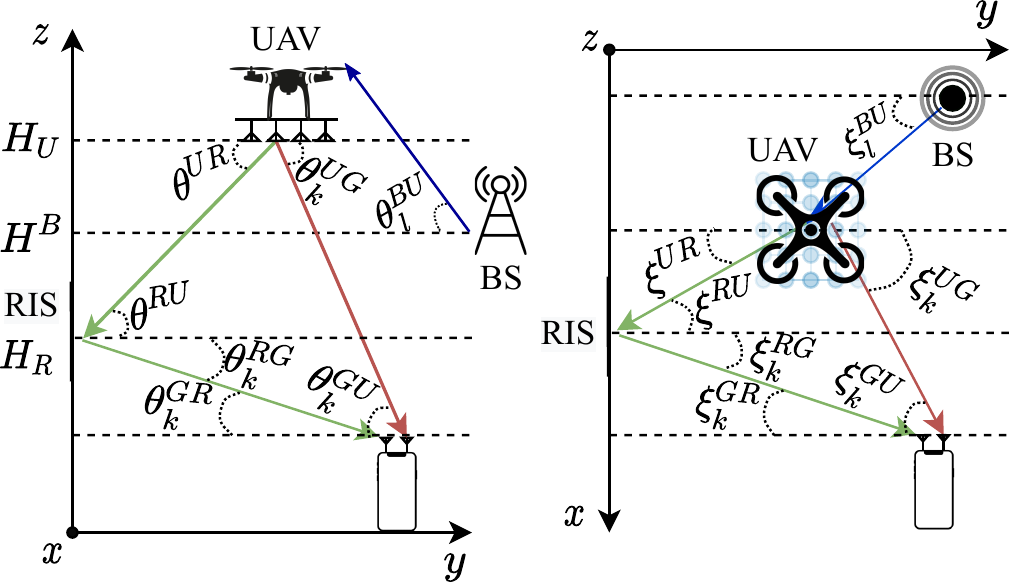}
    \caption{The vertical (left) and horizontal (right) AoDs/AoAs between the \gls{uav}, \gls{ris}, and $k^{th}$ \gls{ue} in the downlink communication system respectively.}
    \label{fig:angles}
\end{figure}
Consider a network environment with $K$ \glspl{ue} randomly spread in the area, a \gls{ris} in fixed and known position. We make the following assumptions:
\begin{itemize}
    \item User association and additional control information needed for data transfer are exchanged between \gls{bs} and \glspl{ue} by means of a dedicated long range control channel.
    \item \glspl{ue} are aware of their own position (e.g., calculated through triangulation with respect to the \glspl{bs} and \glspl{uav}
    in the area).
    \item The \glspl{ue} periodically communicate their position to the \gls{bs} that analyses this information to devise mobility patterns and traffic requirements in the environment.
    \item \glspl{bs}, \glspl{uav} and \glspl{ue} are equipped with \gls{upa} antennas so as to perform concurrent beamforming in different directions.
    \item  An \gls{xlmimo} \gls{ris} deployment is considered in which every \gls{ue} is served by a specific region of the surface. This holds when the \gls{ris} dimensions is large and the \glspl{ue} are sufficiently spaced apart to have partial observability of the surface~\cite{9170651}.
\end{itemize}
\subsubsection{Channel Models}
As visible from \figurename~\ref{fig:problem_scanrio}, there are four channels in the scenario: \gls{bs} to \gls{uav}, \gls{uav} to \gls{ue}, \gls{uav} to \gls{ris} and \gls{ris} to \gls{ue}. The channel gains are denoted as $\mathrm{h}^{BU}$, $\mathrm{h}_k^{UG}$, $\mathrm{h}^{UR}$ and $\mathrm{h}_k^{RG}$ respectively and their derivation is detailed in Appendix I. The \gls{snr} at the \gls{uav} with respect to the associated \gls{bs} is given by,
\begin{equation*} \label{eqn:sinr_u}
    \mathrm{\gamma}[n] = \frac{P^{T}_{BS}{|\mathrm{h}^{\mathrm{BU}}[n]|}^2}{\sigma ^{2}_{n}},
    \tag{1}
\end{equation*}
where $\mathrm{P}_{\gls{bs}}^T$ is the transmit power of the \gls{bs} and $\sigma_n^2$ is the white noise power. Communication using \eqref{eqn:sinr_u} and assuming a Gaussian channel with an \gls{snr} of $\gamma[n]$, the maximum achievable rate of the \gls{bs} to \gls{uav} channel is given by the Shannon bound:
\begin{align*}
    C[n] &= \log_2(1+\gamma[n]) \qquad [\mathrm{bits/s/Hz}].
    \tag{2}
\end{align*}
Note that, to use the above formula, we must assume that both \gls{uav} and  \gls{bs} know the channel between them and can determine the rate based on the available \gls{snr}. Also, we assume that the \glspl{ue} are sufficiently spread apart to avoid mutual interference when communicating with the \gls{uav}. The \gls{snr} at the $k^{th}$ \gls{ue} for the \gls{uav} to \gls{ue} \gls{los} link is given by,
\begin{align*} 
    \mathrm{\gamma}_{k,1} &= \frac{P^{T}_{k}{|\mathrm{h}^{\mathrm{UG}}_k[n]|}^2}{\sigma ^{2}_{n}} \tag{3},\\
\end{align*}
where $P^{T}_{k}$ is the transmit power of the \gls{uav} towards the $k^{th}$ \gls{ue}. Finally, the \gls{snr} at the $k^{th}$ \gls{ue} from the \gls{uav} through the \gls{ris} (forming a cascade channel) is given by,
\begin{align*}
    \mathrm{\gamma}_{k,2} &= \frac{P^{T}_{k}{|\mathrm{H}^{\mathrm{URG}}_k[n]|}^2}{\sigma ^{2}_{n}},
    \tag{4}
\end{align*}
\noindent where $\mathrm{H}^{\mathrm{URG}}_k[n]$ is the overall channel gain of the cascade channel from \gls{uav} to \gls{ris} to \gls{ue}. To be noted that, to limit the number of optimization parameters, we assume that the \gls{uav} transmits with the same power $P^T_k$ on both the direct and indirect (through \gls{ris}) channel to the $k^{th}$ \gls{ue}. Detailed explanation to determine $\mathrm{H}^{\mathrm{URG}}_k[n]$ is given in Appendix I. Assuming, multiple RF chains in the \gls{uav} enough to serve multiple \glspl{ue} in the environment directly and using the \gls{ris}, the total maximum achievable rate in [$bits/s/Hz$] at the $k^{th}$ \gls{ue} from the \gls{uav} can be computed as
\begin{align*}\label{eqn:rate_comb}
    R_{k} = \sum_{i=1}^2 R_{k,i} &= \log_2((1+\gamma_{k,1})(1+\gamma_{k,2})),\\
    \mathrm{where}\\
    R_{k,1} &= \log_2(1+\gamma_{k,1}),\\
    R_{k,2} &= \log_2(1+\gamma_{k,2}).\\
    \tag{5}
\end{align*}
Hence, the scenario can be considered as an alternative to a Spatial Multiplexing scheme where the \gls{ris} redirects the signal from the \gls{uav} to the $k^{th}$ \gls{ue} thereby potentially achieving the rate shown in (\ref{eqn:rate_comb}). 
\subsubsection{Energy Consumption for \gls{uav}} \label{subsec:energy}
\begin{table*}
\centering
\caption{Notation for Energy Consumption Model\cite{cai2020resource}}
\begin{tabular}{|c|c|c|}
\hline
Symbol & Meaning & Simulation Values\\ \hline
$\Omega$ & Blade Angular Velocity & $300~rad/s$ \\ \hline
$r$ & Rotor radius & $0.4~m$ \\ \hline
$\rho$ & Air Density & $1.225~kg/m^3$ \\ \hline
$s$ & Rotor Solidity & $0.05~m^3$ \\ \hline
$A_r$ & Rotor Disc Area & $0.503~m^3$ \\ \hline
$v_0$ & Induced velocity for rotor in forwarding flight & $4.03~m^3$ \\ \hline
$d_0$ & Fuselage drag ratio & $0.3$ \\ \hline
$P_0$ & Blade profile power in hovering status & $79.86~W$ \\ \hline
$P_i$ & Induced power in hovering status & $88.63~W$ \\ \hline
\end{tabular}
\label{tab:energy}
\end{table*}
The power consumption for \gls{uav} is critical due to its limited battery capacity. In the paper, we use the distance-based energy consumption model from~\cite{zeng2019energy  } given by,
\begin{equation*} 
    \mathrm{P}^{UAV}[n]=\underbrace{P_{o}\left(1+\frac{3\Vert \mathbf{v}[n]\Vert^{2}}{\Omega^{2}r^{2}}\right)}_{\mathrm{Bladeprofile}}+\underbrace{\frac{P_{i}v_{0}}{\Vert\mathbf{v}[n]\Vert}}_{\mathrm{Induced}}+\underbrace{\frac{1}{2}d_{0}\rho s A_{\mathrm{r}}\Vert\mathbf{v}[n]\Vert^{3}}_{\mathrm{Parasite}},\tag{6} 
\end{equation*}
where $\mathbf{v}[n]$ is the velocity vector, and the other terms of the equation are explained in Tab.~\ref{tab:energy}. We only considered the energy consumption for the in-flight movement of the \gls{uav} for now and keep the impact of take off and landing on energy consumption for further research.
%Additionally, the \gls{ris} is also an energy limited device that consumes a certain amount of power based on the bit resolution of the elements in \gls{ris}. Typical values for power consumption for the \gls{ris} as mentioned in~\cite{huang2019reconfigurable} are \{1.5, 4.5, 6, 7.8\} mW for \{3-, 4-, 5-, 6-\} bit resolution phase shifting. Hence, the total power consumption for \gls{ris} is given by,
%\begin{align*}
%    \mathrm{P}_{r}^{RIS}[n]=\sum_{k=1}^{K}M_{\mathrm{R}}\mathrm{p}_{r,k}^{RIS}[n]
%    \tag{13}
%\end{align*}
%where $\mathrm{p}_{k,r}^{RIS}[n]$ is the power consumption for one \gls{ris} element of $r^{th}$ \gls{ris} to provide service to $k^{th}$ user at timestep $n$.
\subsubsection{Optimization Problem}
Considering the assumptions, the objective is to find an energy efficient \gls{uav} path and corresponding \gls{ris} phase shift in order to minimize the overall transmission power consumption of \gls{uav} and \gls{bs} under minimum \gls{qos} constraints and maximum \gls{uav} energy budget which is defined as,\\
%where $\mathrm{P}^T_{l,k}[n]$ refers to the total power consumption for data transmission over the (\gls{uav} -\gls{ue}) and (\gls{uav}-\gls{ris}-\gls{ue}) links in timestep $n$ for $l^{th}$ \gls{uav} to their associated \gls{ue}, $\mathrm{P}^{\gls{uav}}[n]$ refers to the power consumption for movement of the $l^{th}$ \gls{uav} in timestep $n$ and $\mathrm{P}^{RIS}_{r}[n]$ refers to the power consumption for each \gls{ris} in timestep $n$. $R_{k,i}$ is the total rate with respect to one path (\gls{uav} to \gls{ue} or \gls{uav} to \gls{ris} to \gls{ue}) and $R_{min}$ is the minimum service rate requirement for the \glspl{ue}. $\mathbf{v}[n]$ and $V_{max}$ are the instantaneous and maximum velocity of the \gls{uav} while  $V_{acc}$ is its acceleration. $\mathrm{C}^{\gls{bs}}$ is the instantaneous capacity achieved for the $l$-th \gls{uav}. $\mathrm{E}^{RIS}_{max}$ refers to the total energy budget of the \gls{ris}. $\tau$ refers to the time duration in each timestep $n$.%
\begin{align*} \label{eqn:opt_prob}
    \min_{\mathbf{P}, \mathbf{Z},\mathbf{V},\mathbf{\Phi}}~ &\sum_{n=1}^N \sum_{k=1}^{K} \mathrm{P}^T_{k}[n] + \sum_{n=1}^N \mathrm{P}^{T}_{BS}[n] \tag{7}\\ 
    s.t.\\
    C1:~ &R_{k}[n] \geq R_{min}, \forall~k,n;\\
    C2:~ &C[n] \geq  \sum_{k=1}^K R_{k}[n],~\forall~n; \\
    C3:~ &0 \leq \mathbf{\Phi}[n] \leq 2\pi;\\
    C4:~ &\sum_{n=1}^N \mathrm{P}^{\gls{uav}}[n] \leq \mathrm{E}^{\gls{uav}}_{max}; \\
    C5:~ &\mathbf{Z}[n+1] = \mathbf{Z}[n] + \mathbf{v}[n]\tau,~n=1,\dots,N-1;\\
    C6:~ &\|\mathbf{v}[n]\| \leq V_{max},~\forall n;\\
    C7:~ &\|\mathbf{v}[n+1]-\mathbf{v}[n]\| \leq V_{acc}\tau,~n=1,\dots,N-1;\\
    C8:~ &\|\mathbf{v}[n]\| \geq 0~\forall n;\\
    C9:~ &\mathbf{Z}[1] = \mathbf{Z}_0;\\
    C10:~ &\mathbf{Z}[N] = \mathbf{Z}_F.\\
\end{align*}
\subsubsection*{Optimization Variables}
The terms of this optimization problem are explained below:
\begin{itemize}
    \item $\mathbf{P}$: \gls{uav} ($\mathrm{P}^T_{k}$) and \gls{bs} ($\mathrm{P}^{T}_{BS}$) transmission power.
    \item $\mathbf{Z}$: \gls{uav} trajectory, represented as the sequence of geographical coordinates of the \gls{uav} at each timestep.
    \item $\mathbf{V}$: \gls{uav} velocity over the trajectory.
    \item $\mathbf{\Phi}$: \gls{ris} phase configurations.
\end{itemize}
\subsubsection*{Objective Function}
The objective is to minimize the overall energy consumption of \gls{uav} and \gls{bs} for transmission during the $N$ timesteps taken by the \gls{uav} to cover its trajectory. 
\subsubsection*{Constraints}
\paragraph*{C1--Guaranteed Rate Constraint}
C1 is devised to provide a guaranteed service rate to each one of the $K$ \glspl{ue}. We recall that $R_{k}$ is the sum rate achieved over the \gls{los} and \gls{ris} link, which has to stay above the guaranteed rate $R_{min}$.
\paragraph*{C2--Backhaul Capacity Constraint}
C2 ensures that the backhaul link capacity is greater than or equal to the aggregate minimum guaranteed rate for all the \glspl{ue}, i.e., that the \gls{uav} has enough bandwidth capacity towards the \gls{bs} to provide at least the minimum guaranteed rate to all the \glspl{ue}.
\paragraph*{C3--Phase Shift Constraint}
C3 limits the phase shift with respect to the incident signal from $0$ to $2\pi$. With the assumption of \gls{xlmimo} surface for \gls{ris}, the phase shift can be considered almost continuous from $0$ to $2\pi$. 
\paragraph*{C4--\gls{uav} Energy Budget}
C4 requires that the total energy consumption of the \gls{uav} over $N$ timesteps does not exceed the threshold $E^{UAV}_{max}$ that defines the maximum energy the \gls{uav} can spend before recharging and, implicitly, the maximum length of the \gls{uav} path.
\paragraph*{C5--Timestep Position Constraint}
C5 constraints the position $\mathrm{Z}[n]$ in successive timesteps, thereby limiting the movement of the \gls{uav} in one timestep.
\paragraph*{C6--Maximum Velocity Constraint}
C6 is devised to constraint the velocity $\mathbf{v}[n]$ of the \gls{uav} in one timestep to be lower than or equal to the maximum velocity $V_{max}$, thereby limiting the maximum distance the \gls{uav} can travel in one timestep.
\paragraph*{C7--Timestep Velocity Constraint}
C7 is devised to determine the velocity $\mathbf{v}[n]$ of the \gls{uav} in successive timesteps based on the maximum acceleration $V_{acc}$ of \gls{uav} in one timestep.
\paragraph*{C8--Minimum Velocity Constraint}
C8 constraints the velocity $\mathbf{v}[n]$ of the \gls{uav}, thereby limiting the minimum distance the \gls{uav} can travel over one timestep. Note that, if the \gls{uav} can hover at one place in one timestep, then the minimum velocity is zero.
\paragraph*{C9/C10--Initial/Final Position Constraint}
C9 and C10 fix the starting and ending points of the trajectory otherwise determined by the optimization problem.

We remark that, as shown in \figurename~\ref{fig:problem_scanrio} the \gls{uav} has two parallel links to each \gls{ue}: one directional and the other with the \gls{ris} sector associated to the \gls{ue}. The multipath approach offers a greater chance to satisfy the service requirement by jointly optimizing the \gls{uav} trajectory $\mathbf{Z}[n]$ and \gls{ris} phase configuration $\mathbf{\Phi}$, while minimizing the transmission power of the entire system. To facilitate the \glspl{ue} to determine the multipath connections, the \gls{bs} has to continuously communicate the beams to be used to the \gls{ue} taking into account the mobility information of the \glspl{ue} and the trajectory of the \gls{uav}. As mentioned before, the \gls{bs} may communicate this information over long-range low-rate technologies such as LoRa \cite{mason2020combining}. 

\section{Analytical Solution}\label{sec:solution}
The optimization problem discussed in the previous section is clearly non-convex and, hence, quite difficult to solve in itself. But we can determine a feasible solution by considering the initial transmission powers for the \gls{uav} and \gls{bs} so as to jointly optimize the \gls{uav} trajectory and \gls{ris} phase and, then, minimize the transmission powers for the given trajectory and phase configuration within the constraints in (\ref{eqn:opt_prob}). This method is explained in detail in the following subsections. 

\subsection{Joint \gls{uav} Trajectory and \gls{ris} phase optimization}
Joint \gls{uav} Trajectory and \gls{ris} phase optimization can be facilitated considering a particular $\mathbf{P}$ over different links~\cite{basar2020simris}. As shown in the \figurename~\ref{fig:angles}, the \gls{bs} to \gls{uav}, \gls{uav} to \gls{ue}, \gls{uav} to \gls{ris} and \gls{ris} to \gls{ue} links are assumed to be deterministic \gls{los} channels. For ease of notation, in the following we indicate the nodes involved in a link using the subscript $U$, $B$, $R$ and $G$ for \gls{uav}, \gls{bs}, \gls{ris} and (ground) \gls{ue}, respectively. The channel information is supposed to be available at the \gls{uav} and the \glspl{ue}.  Hence, to maximise the transmission efficiency, a \gls{mrt} is applied, i.e., the transmission beamformer for any $k^{th}$ \gls{ue} as well as for the \gls{uav} can be defined as $\mathrm{w^{BU}} = \frac{1}{\sqrt{M_B}}\mathrm{h^{BU}}, \mathrm{w^{UG}_{k}} = \frac{1}{\sqrt{M_U}}\mathrm{h^{UG}}$ and $\mathrm{w^{UR}} = \frac{1}{\sqrt{M_U}}\mathrm{h^{UR}}$. The overall channel gains can hence be obtained as,
\begin{align*} 
    \mathrm{H^{BU}}[n]&=(\mathrm{h^{BU}}[n])^{\mathrm{H}}[n]\mathrm{w^{BU}}[n]=\frac{\sqrt{M_{\mathrm{B}}}\alpha_{0}}{d^{\mathrm{BU}}[n]}; \tag{8}\\
    \mathrm{H^{UG}_{k}}[n]&=(\mathrm{h^{UG}_{k}}[n])^{\mathrm{H}}[n]\mathrm{w^{UG}_{k}}[n]=\frac{\sqrt{M_{\mathrm{U}}}\alpha_{0}}{d_{k}^{\mathrm{UG}}[n]};\ \tag{9}\\ \mathrm{H^{URG}_{k}}[n]&=(\mathrm{h_{k}^{\mathrm{RG}}}[n])^{\mathrm{H}}\mathbf{\Phi}_{k}[n]\mathrm{H}^{\mathrm{UR}}[n]\mathrm{w^{UR}}[n]\\ &=\sqrt{M_{\mathrm{U}}}(\mathrm{h_{k}^{\mathrm{RG}}}[n])^{\mathrm{H}}\mathbf{\Phi}_{k}[n]\mathrm{h^{\mathrm{RU}}}[n]\\
    &=\frac{\sqrt{M_{\mathrm{U}}}M_{\mathrm{R}}\alpha_{0}}{d_{k}^{\mathrm{RG}}d^{\mathrm{UR}}[n]}.\tag{10} 
\end{align*}
To determine the $\mathrm{H^{URG}_{k}}[n]$ coefficients correctly, the optimal phase control policy for the phase shift in every timestep (which maximizes the reflection-mode channel gain by aligning the phase of the \gls{ris} to match those of the channel) is given by,
\begin{align*} 
    &\mathbf{\Phi}_{m_{\mathrm{R}x}, m_{\mathrm{R}y},k}=\frac{2\pi\Delta_{\mathrm{R}}}{\lambda_{\mathrm{c}}}[(m_{\mathrm{R}x}-1)(\sin\theta^{\mathrm{RU}}\cos\xi^{\mathrm{RU}}\\ &+\sin\theta_{k}^{\mathrm{RG}}\cos\xi_{k}^{\mathrm{RG}})+(m_{\mathrm{R}y}-1)(\sin\theta^{\mathrm{RU}}\sin\xi^{\mathrm{RU}}\\ &+\sin\theta_{k}^{\mathrm{RG}}\sin\xi_{k}^{\mathrm{RG}})], 
    \tag{11}
\end{align*}
where $\theta^{\mathrm{RU}}$ and $\xi^{\mathrm{RU}}$ are the \glspl{aoa} and $\theta^{\mathrm{RG}}$ and $\xi^{\mathrm{RG}}$ are the \glspl{aod} as defined in \figurename~\ref{fig:angles}. The assumption for \gls{ris} phase configuration is that there is a wired direct link to the \gls{ris} controller and that, delay and imperfect phase configuration are negligible. \\
Note that, the problem is still non-convex due to C1 and C2 w.r.t. $\mathbf{Z}$. In order to overcome this issue, we add three slack variables $\lambda_{k,i}[n]$, $\mu[n]$ and $\pi[n]$. In this way, we keep the constraints C1-9, and the problem can be reformulated as follows,
\begin{align*} \label{eqn:opt_prob_reform}
    \min_{\mathbf{Z},\mathbf{V},\mathbf{\Lambda},\mathbf{M},\mathbf{\Pi}}~& \sum_{k=1}^K \sum_{n=1}^N \mathrm{P}^T_{k}[n]  + \sum_{n=1}^N \mathrm{P}^{T}_{BS}[n] \tag{12}\\ 
    s.t.\\
    C1 - ~& C10;\\
    C11:~ &\|\mathbf{Z}_k^{\gls{ue}} - \mathbf{Z}[n]\|^2 \leq \lambda_{k,1}[n], k \in \{1, ..., K\}; \\
    C12:~ &\|\hat{\mathbf{Z}}_k^{\gls{ris}} - \mathbf{Z}[n]\|^2 \leq \lambda_{k,2}[n], k \in \{1, ..., K\};\\
    C13:~ &\|\mathbf{Z}^{\gls{bs}} - \mathbf{Z}[n]\|^2 \leq \mu[n];\\
    C14:~ &\|\mathbf{v}[n]\|^2 \geq \pi^2[n];\\
    C15:~ & \pi[n] \geq 0;
\end{align*}
where $\mathbf{\Lambda} = \{\lambda_{k,i}[n], \forall~n,k,i\}$, $\mathbf{M} = \{\mu[n], \forall~n,l\}$ and $\mathbf{\Pi} = \{\pi[n], \forall~n\}$.  
Similarly to what proposed in~\cite{cai2020resource}, we overcome the non-convex constraints C1 and C2 via \gls{sca} in an iterative way. We can compute a lower bound of the instant achievable rate for each user by modifying $\lambda_{k,i}[n]$, $\mu[n]$ and $\pi[n]$ and calculating the first-order Taylor expansion which is a global under-estimator of the rate convex function~\cite{zeng2017energy}. Hence, omitting the argument $[n]$ for notation clarity, we redefine the \gls{snr} expression as,
\begin{align*}
    \gamma_{k,i} &= \frac{\hat{\gamma}_{k, i}}{\lambda_{k, i}}, i=1, 2; \tag{13} \\
    \gamma &= \frac{\hat{\gamma}}{\mu};
    \tag{14}
\end{align*}
where
\begin{align*}
    \hat{\gamma} &= \frac{\mathrm{P}^{T}_{BS}\mathrm{M_B}\alpha^2_0}{\sigma ^{2}_{n}}; \tag{15}\\
    \hat{\gamma}_{k,1} &= \frac{\mathrm{P}^{T}_{k}\mathrm{M_U}\alpha^2_0}{\sigma ^{2}_{n}}; \tag{16}\\
    \hat{\gamma}_{k,2} &= \frac {\mathrm{P}^{T}_{k}\mathrm{M_U}\mathrm{M^2_R}\alpha^2_0}{(d^{\mathrm{RG}}_k)^2\sigma ^{2}_{n}}.
    \tag{17}\label{eq:snr_ris}
\end{align*}
Hence, the maximum achievable instant rate per link $i$ for $k^{th}$ \gls{ue} from the \gls{uav} is given by,
\begin{align*}
    \hat{R}_{k,i}[n] &= \log_2(1+\gamma_{k, i}[n]), \label{eqn:rate_1}\tag{18}
\end{align*}
Similarly, the capacity at the \gls{uav} from \gls{bs} is given by,
\begin{align*}
    \hat{C}[n] &= \log_2(1+\gamma[n])\label{eqn:rate_2}.
    \tag{19}
\end{align*}
Applying the first-order Taylor expansions in the $j$-th iteration for a particular value of $\lambda_{k, i}^j[n]$, $\mu^j[n]$ and $\mathbf{v}^j[n]$ in (\ref{eqn:rate_1}) and (\ref{eqn:rate_2}), the lower bound for the rates is given by,
\begin{align*} 
    \hat{R}_{k, i}[n] \geq &\\(\hat{R}_{k, i}[n])^{j}=&~\log_{2}\left(1+\frac{\gamma_{k, i}[n]}{\lambda_{k, i}^{j}[n]}\right)\\ &-\frac{\gamma_{k, i}[n](\lambda_{k, i}[n]-\lambda_{k, i}^{j}[n])}{\lambda_{k, i}^{j}[n](\lambda_{k, i}^{j}[n]+\gamma_{k, i}[n])\ln 2},
    \tag{20}\label{eqn:rate}\\
    \hat{C}[n] \geq &\\(\hat{C}[n])^{j} =&~\log_{2}\left(1+\frac{\gamma[n]}{\mu^{j}[n]}\right)\\ -&\frac{\gamma[n](\mu[n]-\mu^{j}[n])}{\mu^{j}[n](\mu^{j}[n]+\gamma[n])\ln 2},
    \tag{21}\label{eqn:capacity}\\
    \Vert \mathbf{v}[n]\Vert^{2} \geq &~\Vert \mathbf{v}^j[n]\Vert^{2}+2[\mathbf{v}^j[n]]^{\mathrm{T}}(\mathbf{v}[n]-\mathbf{v}^j[n]).\tag{22}\label{eqn:velocity}
\end{align*}
where $(\hat{R}_{k, i}[n])^{j}$ and $(\hat{C}[n])^{j}$ are the lower bound achievable rates for the $k^{th}$ \gls{ue} and \gls{uav} respectively, in the $j^{th}$ iteration of \gls{sca}.\\
Additionally, the total transmission energy consumed over the whole trajectory can be represented as,
\begin{align*}
    \mathrm{P}^{Total} = &\sum_{k=1}^K \sum_{n=1}^N \mathrm{P}^T_{k}[n] + \sum_{n=1}^N \mathrm{P}^{T}_{BS}[n], \tag{23} \\
\end{align*}
Also, the in-flight power consumption for \gls{uav} can be written as,
\begin{align*}
    \mathrm{P}^{UAV}[n]&=P_{o}\left(1+\frac{3\Vert \mathbf{v}[n]\Vert^{2}}{\Omega^{2}r^{2}}\right)+\frac{P_{i}v_{0}}{\pi[n]}+\frac{1}{2}d_{0}\rho_{s} A_{\mathrm{r}}\Vert \mathbf{v}[n]\Vert^{3}.
    \tag{24}
\end{align*}
Applying the lower bounds in (\ref{eqn:rate}), (\ref{eqn:capacity}) and (\ref{eqn:velocity}) in (\ref{eqn:opt_prob_reform}) we obtain a convex problem defined as,
\begin{align*} \label{eqn:opt_prob_reform_sca}
    \min_{\mathbf{Z},\mathbf{V},\mathbf{\Lambda},\mathbf{M},\mathbf{\Pi}} &~\mathrm{P}^{Total} \tag{25}\\ 
    s.t.\\
    \hat{C}1:&~\sum_{i=1}^2 (\hat{R}_{k, i}[n])^{j} \geq R_{min}, \forall~k;\\
    \hat{C}2:&~(\hat{C}[n])^{j} \geq \sum_{k=1}^K\sum_{i=1}^2 (\hat{R}_{k, i})^{j}[n]; \\
    \hat{C}14:&~\Vert\mathbf{v}^j[n]\Vert^{2}+2[\mathbf{v}^j[n]]^{\mathrm{T}}(\mathbf{v}[n]-\mathbf{v}^j[n]) \geq \pi^2[n];\\
    C3&-C15,
\end{align*}
which solving it provides an upper bound of the problem in (\ref{eqn:opt_prob_reform}). We iteratively update the feasible solution $\mathbf{Z}^j[n]$, $\lambda_{k,i}^j[n]$, $\mu^j[n], \mathbf{v}^j[n]$ and $\pi^j[n]$ by solving the convex problem in (\ref{eqn:opt_prob_reform_sca}) using the CVX standard optimization solver~\cite{grant2009cvx} in the $j$-th iteration. 
\subsection{Transmission Power Control}
For a determined \gls{uav} trajectory and \gls{ris} phase, the \gls{uav} and \gls{bs} transmission power can be minimized. To define the transmission power minimization with a predefined trajectory $\mathbf{Z}$, the optimization problem in (\ref{eqn:opt_prob}) can be rewritten with constraints $C3-C10$ already satisfied for the pre-defined trajectory $\mathbf{Z}$. So the optimization problem can be written as,
\begin{align*}\label{eqn:opt_prob_power}
   \min_{\mathbf{P}}~ &\mathrm{P}^{Total} \tag{26}\\
    s.t.\\
    C1:~ &\sum_{i=1}^2 R_{k,i}[n] \geq R_{min}, \forall~k,n;\\
    C2:~ &C[n] \geq \sum_{k=1}^K\sum_{i=1}^2 R_{k, i}[n],~\forall~n. \\
\end{align*}
The constraints $C1$ and $C2$ are concave with respect to the $\mathbf{P}$. Hence it can be easily solved by employing \gls{sca} using the Taylor's expansions of (\ref{eqn:rate_1}) and (\ref{eqn:rate_2}), which are global over-estimators of the concave functions. To do so, the \gls{snr} expressions are rewritten as,
\begin{align*}
    \gamma_{k,i}[n] &= \mathrm{P}^T_{k, i}[n]\kappa_{k,i}[n], \forall~i\in\{1,2\},~k; \tag{27}\\
    \gamma[n] &= \mathrm{P}^T_{BS}[n]\kappa[n]; 
    \tag{28}
\end{align*}
where,
\begin{align*}
    \kappa_{k,1}[n] &= \bigg\{\frac{\sqrt{M_{\mathrm{U}}}\alpha_{0}}{ d_{k}^{\mathrm{UG}}[n]\sigma}\bigg\}^2, \tag{29}\\
    \kappa_{k,2}[n] &= \bigg\{\frac{\sqrt{M_{\mathrm{U}}}M_{\mathrm{R}}\alpha_{0}}{d_{k}^{\mathrm{RG}}d^{\mathrm{UR}}[n]\sigma}\bigg\}^2, \tag{30} \\
    \kappa[n] &= \bigg\{\frac{\sqrt{M_{\mathrm{B}}}\alpha_{0}}{ d^{\mathrm{BU}}[n]\sigma}\bigg\}^2. \tag{31} \\
\end{align*}
The first-order Taylor expansion for (\ref{eqn:rate_1}) and (\ref{eqn:rate_2}) yields,
\begin{align*}
    \hat{R}_{k, 1}[n] \leq &\\(\hat{R}_{k, 2}[n])^{j}=&~\log_{2}\left(1+(\mathrm{P}^T_{k, 1})^j{|\kappa_{k,1}[n]|}^2\right)\\&+\frac{{|\kappa_{k,1}[n]|}^2(\mathrm{P}^T_{k, 1}-(\mathrm{P}^T_{k, 1})^j)}{(1+(\mathrm{P}^T_{k, 1})^j{|\kappa_{k,1}[n]|}^2)\ln(2)}, \tag{32}\\
    \hat{R}_{k, 2}[n] \leq &\\(\hat{R}_{k, 2}[n])^{j}=&~\log_{2}\left(1+(\mathrm{P}^T_{k, 2})^j{|\kappa_{k,2}[n]|}^2\right)\\&+\frac{{|\kappa_{k,2}[n]|}^2(\mathrm{P}^T_{k, 2}-(\mathrm{P}^T_{k, 2})^j)}{(1+(\mathrm{P}^T_{k, 2})^j{|\kappa_{k,2}[n]|}^2)\ln(2)}, \tag{33}\\
    \hat{C}[n] \leq &\\(\hat{C}[n])^{j} =&~\log_{2}\left(1+(\mathrm{P}^T_{BS})^j{|\kappa[n]|}^2\right)\\&+\frac{{|\kappa_{l}[n]|}^2(\mathrm{P}^T_{BS}-(\mathrm{P}^T_{BS})^j)}{(1+(\mathrm{P}^T_{BS})^j{|\kappa[n]|}^2)\ln(2)}, \tag{34}
\end{align*}
Hence, the optimization problem (\ref{eqn:opt_prob_power}) can be rewritten as,
\begin{align*}\label{eqn:opt_prob_power_2}
   \min_{\mathbf{P}}~ &\mathrm{P}^{Total} \tag{35}\\
    s.t.\\
    C1:~ &\sum_{i=1}^2 (\hat{R}_{k, i}[n])^{j}[n] \geq R_{min}, \forall~k,n;\\
    C2:~ &(\hat{C}[n])^{j}[n] \geq  \sum_{k=1}^K\sum_{i=1}^2 (\hat{R}_{k, i})^{j}[n],~\forall~n. \\
\end{align*}
Similar to the \gls{uav} trajectory and \gls{ris} phase optimization problem, this optimization can be solved using the CVX standard optimization solver. Algorithm \ref{algo:joint} provides the pseudocode to solve the optimization problem. Note that, we are aiming to minimize the transmission energy consumption of the \gls{uav} and \gls{bs} by iteratively choosing a \gls{uav} route aided by the phase shifting involved in the \gls{ris}, which satisfies a target minimum rate for all the users, taking into account the rate aggregation of all the users is achievable, fulfilling the backhaul link capacity limitation, something not addressed in the literature to the best of the authors' knowledge.\\

\begin{algorithm}[t!]
\SetAlgoLined
\KwResult{\gls{uav} Trajectory $\mathbf{Z}$, \gls{uav} Velocity $\mathbf{V}$, $\mathrm{P}^{Total}$}
Initialize trajectory $\mathbf{Z}$, $\mathbf{V}$, maximum number of iteration $J_{max}$, initial iteration index $j=0$, Particular UAV and BS transmission power $\mathbf{P}$ , Initial trajectory $\mathbf{Z}$, Initial velocity $\mathbf{V}$   and Convergence tolerance $\epsilon$\;
\While{$j~\leq~J_{max}$ or $\dfrac{\mathrm{P}_{Total}^{j} - \mathrm{P}_{Total}^{j-1}}{\mathrm{P}_{Total}^j} \leq \epsilon$ } {
Set $j = j + 1$ and $\{\mathbf{P}^j,\mathbf{V}^j,\mathbf{\Lambda}^j,\mathbf{M}^j, \mathbf{\Pi}^j\} = \{\mathbf{P},\mathbf{V},\mathbf{\Lambda},\mathbf{M}, \mathbf{\Pi}\}$\;
Solving optimization problem (\ref{eqn:opt_prob_reform_sca}) to obtain $\mathbf{Z}, \mathbf{V}, \mathbf{\Lambda}, \mathbf{M} $ and $\mathbf{\Pi}$ for a Particular $\mathbf{P}$\;
Solving optimization problem (\ref{eqn:opt_prob_power_2}) to obtain $\mathbf{P}$ and $\mathrm{P}_{Total}$ for a Particular $\mathbf{Z}, \mathbf{V}, \mathbf{\Lambda}, \mathbf{M}, \mathbf{\Pi}$\\
Update $\mathrm{P}_{Total}^j = \mathrm{P}_{Total}$\; 
}
\caption{Joint Trajectory, RIS Phase Configuration and Transmission Power Control algorithm}
\label{algo:joint}
\end{algorithm}

\section{Results and Discussion}\label{sec:results}
The solution discussed in the previous section is implemented in MATLAB simulation environment. The base simulation parameters are defined by Tab. \ref{tab:simulation}. 
\subsection{Simulation Environment}
\begin{figure}[t]
    \centering
    \includegraphics[width=0.45\textwidth]{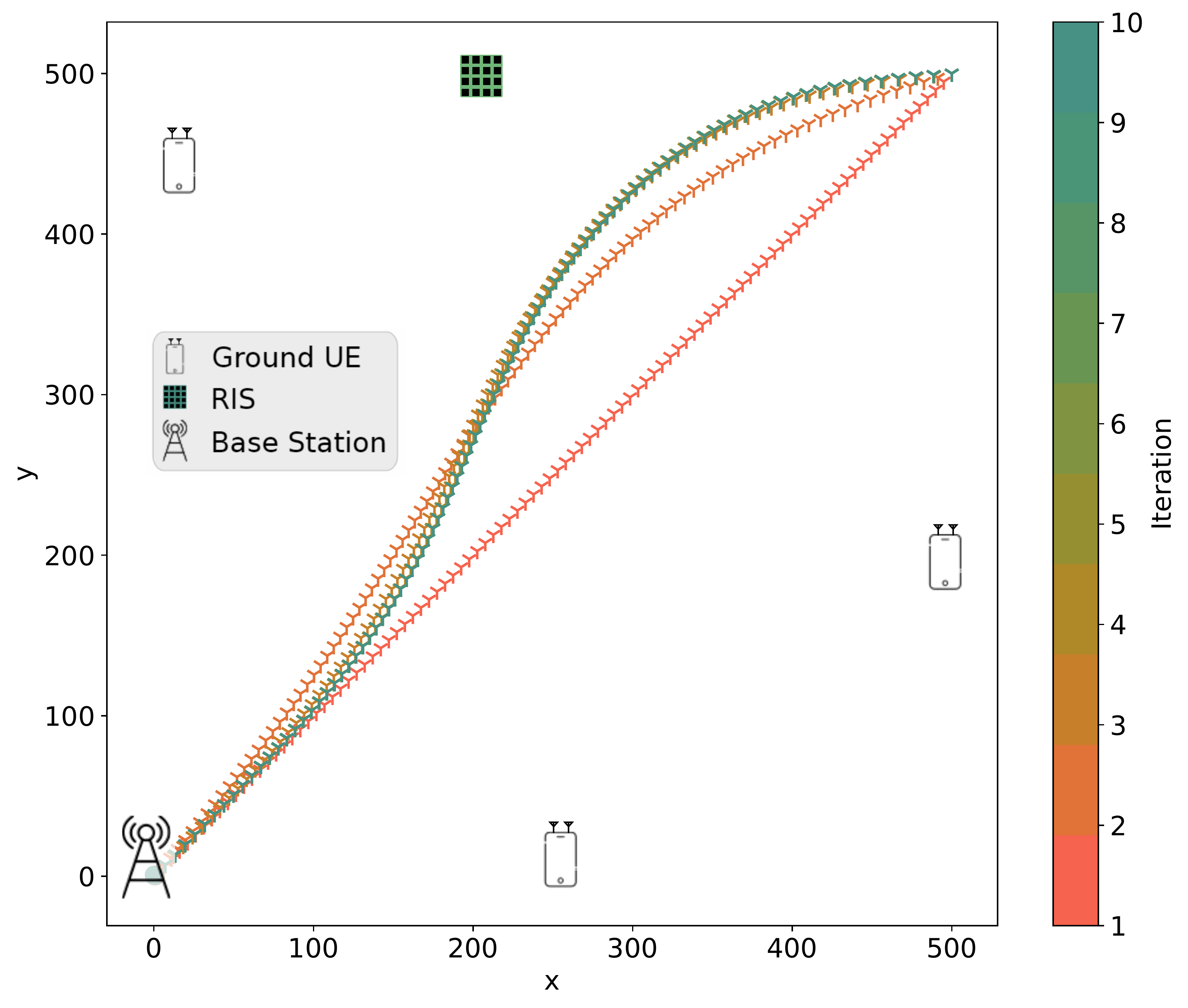}
    \caption{Optimization of the \gls{uav} trajectory and transmission power over \gls{sca} iterations. The marked lines represent the \gls{uav} trajectories obtained during the execution of the iterative algorithm. The straight line is the initial solution, while the darkest one is the final solution.}
    \label{fig:iterations}
\end{figure}

\begin{table}[t!]
\caption{Simulation Parameters}
\centering
\begin{tabular}{|c|c|}
\hline
Parameter                      & Value                    \\ \hline
Area                           & 500m $\times$ 500m                 \\ \hline
Number of Users (K)               & 3                        \\ \hline
Position of Users                & [20, 450; 250, 0; 500, 200]                        \\ \hline
Position of Base Station                & [0, 0]                        \\ \hline
Number of UAVs                 & 1                        \\ \hline
Initial/Final Position of UAV ($\mathrm{Z}_0/\mathrm{Z}_F$)                & [0, 0; 500, 500]                        \\ \hline

Maximum Velocity               & 20 m/s                   \\ \hline
Maximum Acceleration           & 4 m/s$^2$ \\ \hline
Height of the [UAV, BS, RIS] & [20,15,10] m           \\ \hline
Path Loss ($\alpha_0$)                      & 61 dBm        \\ \hline
Noise Power Spectral Density ($\sigma^2$)   & -174 dBm                 \\ \hline
\end{tabular}
\label{tab:simulation}
\end{table}
At this stage of the work, we only considered scenarios with static \glspl{ue}. The analysis of the system performance in presence of mobile \glspl{ue} is left to future work. The simulation results are categorised under five different evaluation scenarios:
\begin{itemize}
    \item \emph{\gls{uav} \gls{los} transmission power v/s \gls{uav} \gls{ris} transmission power}: The evaluation scenario shows the transmission power necessary to be used using the \gls{los} link and the \gls{ris} link.
    \item \emph{Impact of \gls{ris} Position on \gls{uav} Trajectory}: The scenario studies the impact of the \gls{ris} position on the \gls{uav} trajectory.
    \item \emph{Impact of \gls{ris} Position on \gls{uav} \gls{los} Transmission Power}: The scenario analyzes the impact of the \gls{ris} position on the \gls{los} transmission power consumption when vary the minimum rate requirements.
    \item \emph{Impact of \gls{ris} Position on \gls{uav} Trajectory Power Consumption}: The scenario summarises the \gls{uav} trajectory power consumption for different \gls{ue} minimum rate requirements. 
    \item \emph{Impact of \gls{uav} Energy Budget on \gls{uav} Trajectory}: The scenarios shows the impact of the total energy budget $E^{UAV}_{max}$ for the \gls{uav} trajectory.  
\end{itemize}
The simulation environment (with parameters denoted in Tab. \ref{tab:simulation}) is shown in the \figurename~\ref{fig:iterations}. As visible from the figure, over the \gls{sca} iterations, the \gls{uav} trajectory and transmission power is optimized using Algorithm \ref{algo:joint} until it converges, i.e., \gls{uav} trajectory and transmission power are no longer improved. 

\begin{figure}[t]
    \centering
    \includegraphics[width=0.45\textwidth]{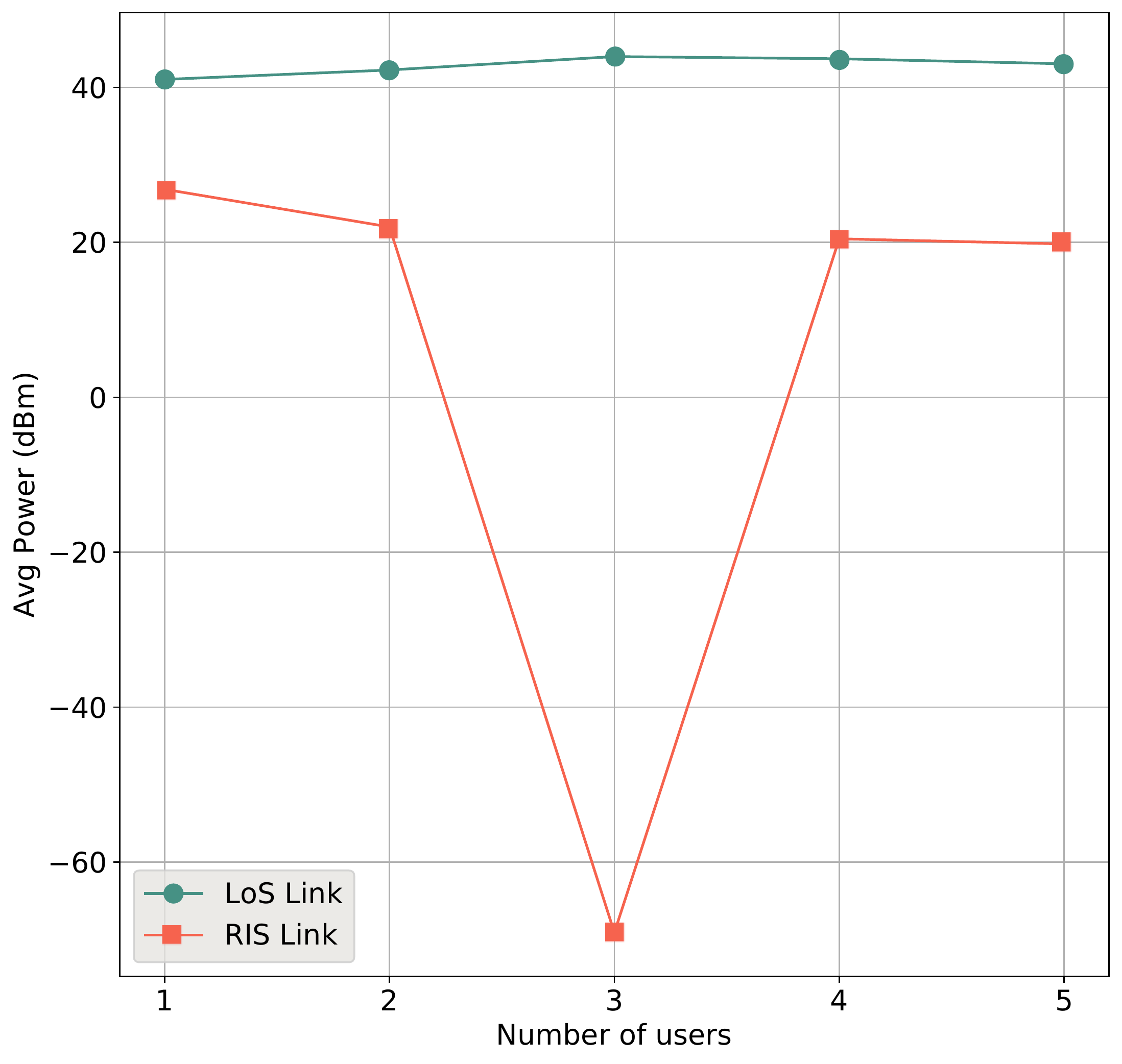}
    \caption{Average Power Consumption over \gls{uav} \gls{los} and \gls{ris} links towards the \glspl{ue}, for increasing number of static \glspl{ue} and with a constant minimum rate requirement $R_{min}$.}
    \label{fig:avg_power}
\end{figure}
\subsection{\gls{uav} \gls{los} Transmission Power v/s \gls{uav} \gls{ris} Transmission Power}
\begin{figure}[t]
    \centering
    \includegraphics[width=0.45\textwidth]{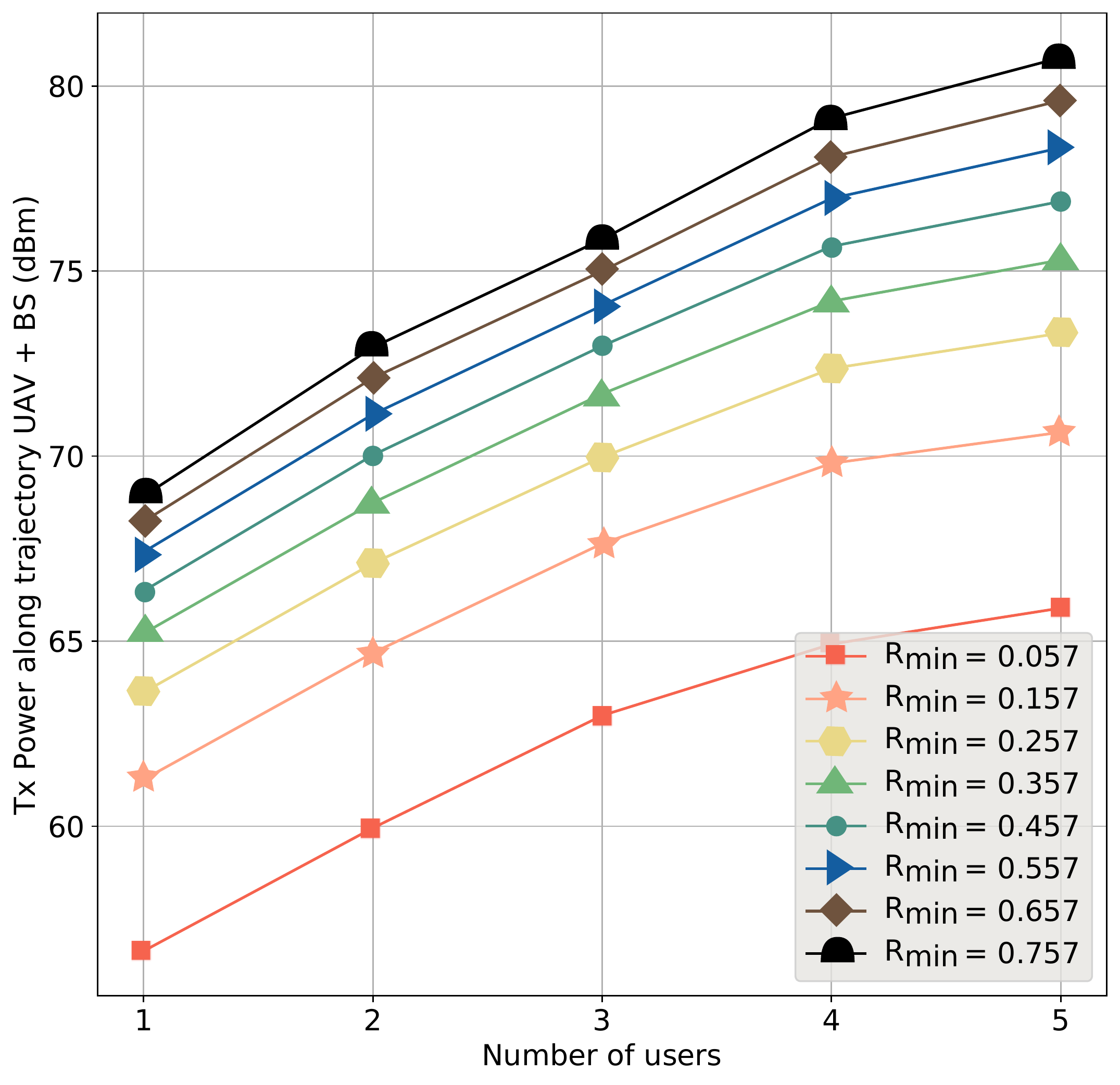}
    \caption{Total Transmission Power (\gls{uav} and \gls{bs}) when increasing the number $K$ of static \glspl{ue} for different values of $R_{min}$.}
    \label{fig:tx_power_uav_bs}
\end{figure}

\begin{figure}[t]
    \centering
    \includegraphics[width=0.45\textwidth]{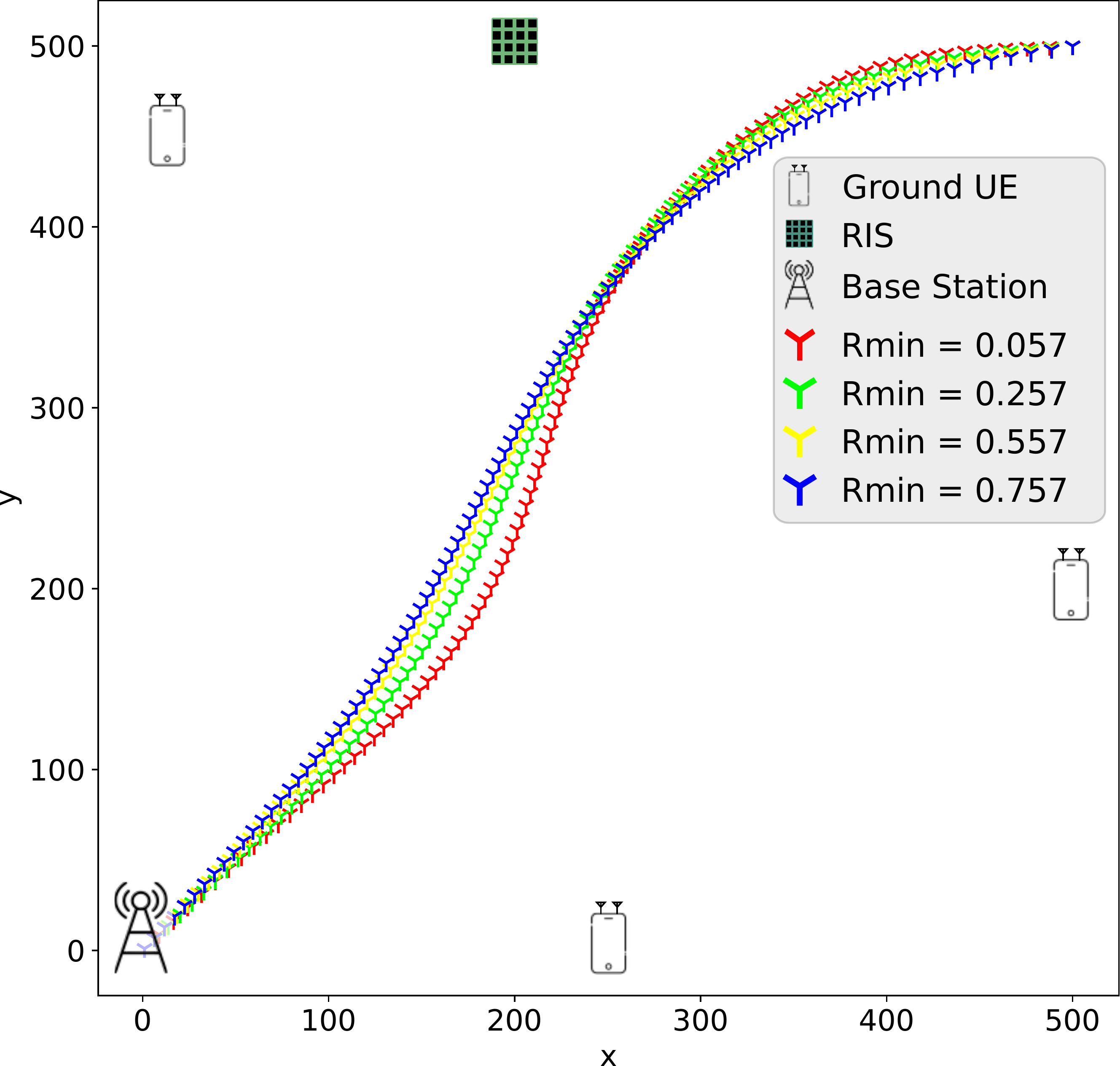}
    \caption{Different Optimal Trajectories for the \gls{uav} for three \glspl{ue} and for different values of $R_{min}$.}
    \label{fig:uav_tr}
\end{figure}

Different configurations in terms of static number of \glspl{ue} in the network have been simulated. \figurename~\ref{fig:avg_power} shows the average transmission power per timestep along the optimized trajectory for both the \gls{los} and the \gls{ris} links for $K$ \glspl{ue}. The first observation is the transmission power over \gls{ris} link is significantly lower than that over \gls{los}. This shows the fundamental role of the $M_R^2$ factor in (\ref{eq:snr_ris}) to help provide good \gls{snr} conditions. The results show that, in general, the average transmission power over the \gls{los} link is slightly increasing for $K\leq3$ and then slightly decreasing for $K>3$. Also, there are significant changes in the \gls{ris} link. It is noticed that the transmission power for \gls{los} link (bullet-marked) generally increases for $K\leq3$ \glspl{ue}, i.e., \gls{los} link is preferred, while, when $K>3$, the \gls{ris} link is preferred reducing the \gls{los} contribution to fulfill the constraints while increasing in the \gls{ris} link transmission power. Additionally, the sudden drop in average power consumption for \gls{ris} link for three \glspl{ue} is because the \gls{ris} is far away from the \glspl{ue} as visible from \figurename~\ref{fig:iterations}. Hence, the system is very sensitive to the \gls{ris} position. The variations in the \gls{ris} link also shows the importance of its usage, as it adapts to the environment providing less or more power in order to fulfill the constraints. This shows the potential impact of \gls{ris} in terms of total power minimization and scalability of the system. Looking at the total transmission power used along the optimized trajectory, that is, the summation of the power from the \gls{bs} and the transmission power for the \gls{uav}, \figurename~\ref{fig:tx_power_uav_bs} shows now that the power increases with the number of users in the network for different values of $R_{min}$. To be noted that the curve bends when the number of users increases, since their distance to the \gls{bs}, \gls{uav} and \gls{ris} reduces. Also, the total power increases with the minimum rate requirement. Another significant observation is the change in average power consumption per set of users for different rates. The change in principle should be exponential, i.e., linear increase in rate should require exponential increase in power. But, to follow this criteria, the distance has to be constant, i.e., the trajectory of the \gls{uav} has to be constant for all the different rates. But, as visible in \figurename~\ref{fig:uav_tr}, which shows the optimal trajectories for different values of $R_{min}$, the optimal trajectory for $R_{min} = 0.057$ is able to deviate more from the straight line trajectory as it can still satisfy the low required minimum rate for the \glspl{ue}. On the contrary, the optimal trajectory for $R_{min} = 0.757$ is able to deviate less from the straight line trajectory than that for $R_{min} = 0.057$ as the required minimum rate is higher. Note that, we only show optimal trajectories for $R_{min} = \{0.057,0.257,0.557,0.757\}$ to be able to visually distinguish between the optimal trajectories for the different values of $R_{min}$. The optimal trajectories for the remaining values of $R_{min}$ are between the optimal trajectory for $R_{min} = 0.057$ and $R_{min} = 0.757$. This trend for optimal trajectories is also true for the scenarios involving one, two, four and five \glspl{ue}. Hence, the average power consumption for \glspl{ue}, as shown in \figurename~\ref{fig:tx_power_uav_bs}, does not follow an exponential criteria due to change in optimal trajectory for different values of $R_{min}$. Additionally, the system fails to find feasible solutions above five \glspl{ue}, i.e., one \gls{uav} cannot serve more than five \glspl{ue} simultaneously in a single flight in the considered scenario. The current configuration based on CVX, makes difficult to go beyond $R_{min} = 0.757$. Then, the usage of reinforcement learning can be explored to improve scalability, which has been left for future work.   
\subsection{Impact of \gls{ris} Position on \gls{uav} Trajectory}
\begin{figure}[t]
    \centering
    \includegraphics[width=0.45\textwidth]{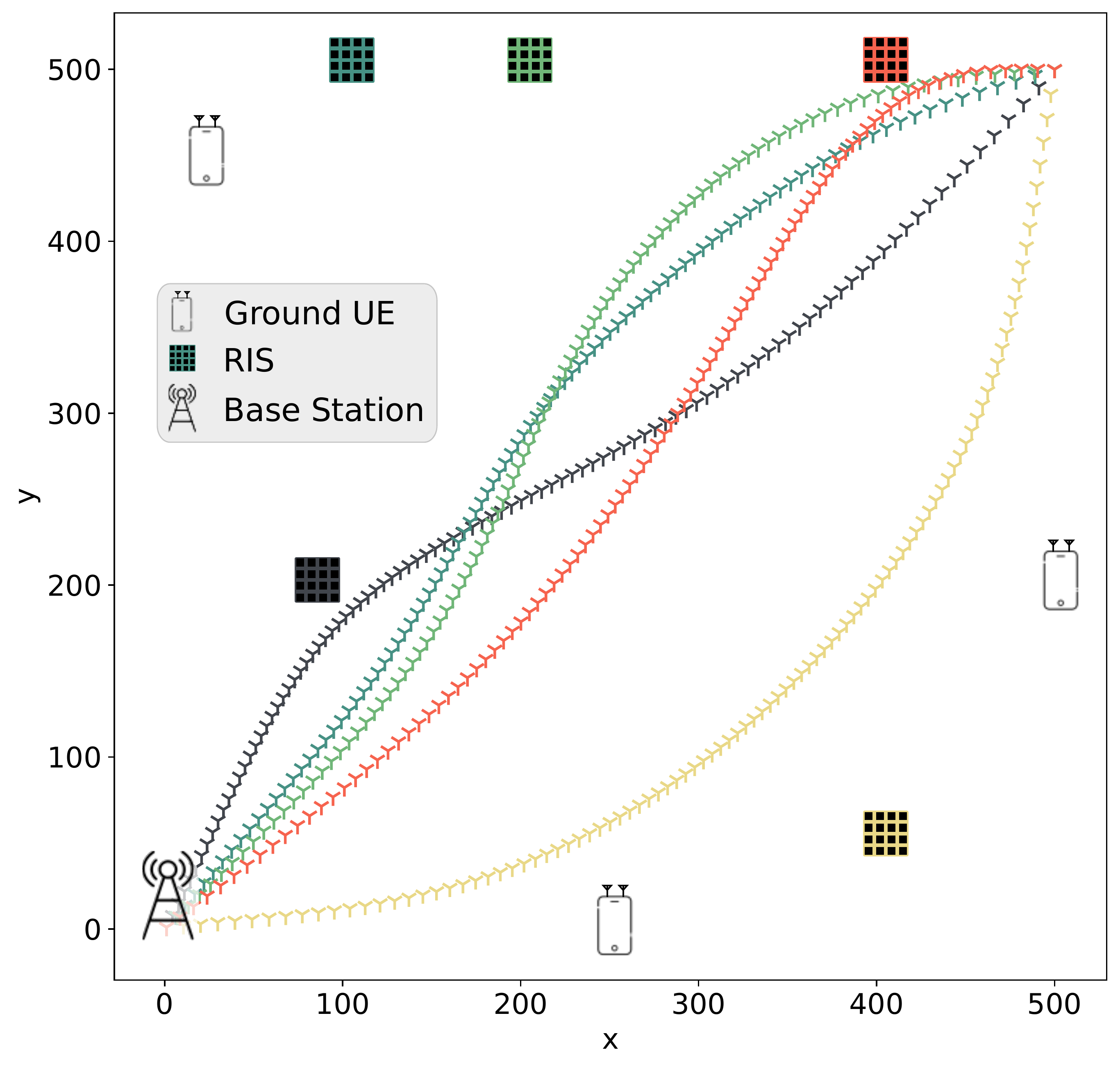}
    \caption{Impact of \gls{ris} position on \gls{uav} Trajectory for a fixed \gls{ue} and \gls{bs} positions}
    \label{fig:ris_pos}
\end{figure}
We analyze the impact of the different positions of \gls{ris} on the optimal \gls{uav} trajectories as shown in \figurename~\ref{fig:ris_pos}. The first discernible observation is that the \gls{uav} attempts to go as close as possible to the \gls{ris}. This is because the transmission power necessary to satisfy the rate requirement of the \glspl{ue} is considerably lower when using the \gls{ris} compared to directly transmitting to the \glspl{ue}. This, however, is compensated by the higher path loss that is encountered by the signal i.e. the total distance the signal has to cover using the \gls{ris} is higher than that along the \gls{los} channel. Due to this fact, the \gls{uav} cannot use the \gls{ris} to serve all the \glspl{ue} at all timesteps. Hence some \glspl{ue} has to be served directly. This creates a push-pull effect on the \gls{uav} that prevents the \gls{uav} to venture very close to one \gls{ue} to avoid violating the QoE requirements of the other \glspl{ue}. Hence, determining an optimal position for \gls{ris} is important while designing the network.
\subsection{Impact of \gls{uav} Energy Budget on \gls{uav} Trajectory}
\begin{figure}[t!]
    \centering
    \includegraphics[width=0.45\textwidth]{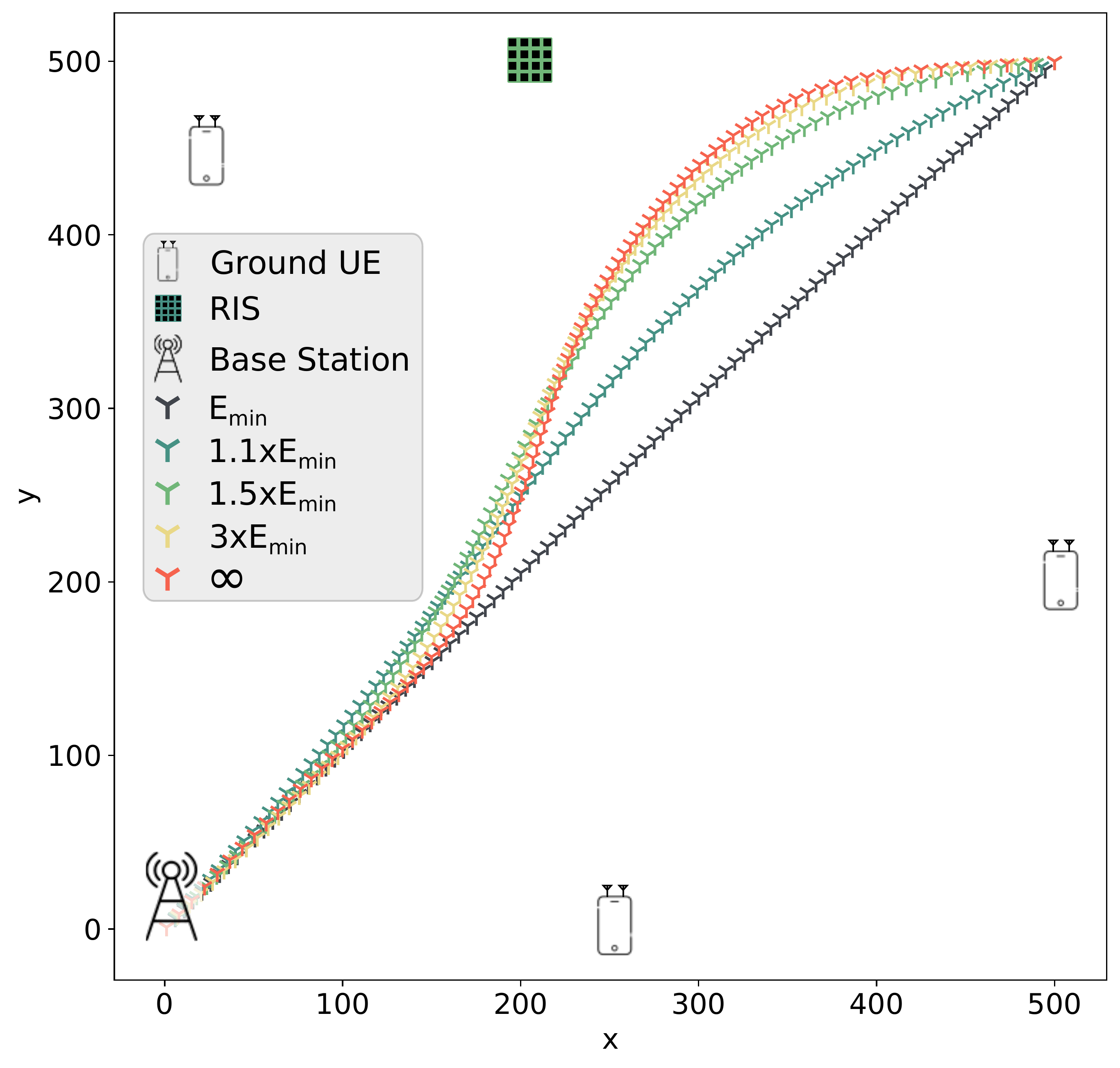}
    \caption{Impact of \gls{uav} Energy Budget on the \gls{uav} Trajectory}
    \label{fig:uav_energy}
\end{figure}
To study the impact of the energy budget on the \gls{uav} trajectory optimization, we determine the \gls{uav} trajectories for different energy budget values. The energy consumed by the \gls{uav} over the straight line path (i.e, shortest path) is the minimum in-flight energy consumption necessary for the \gls{uav} to reach its final destination and is hence set as a reference minimum $\mathrm{E_{min}}$. Hence, the energy budget for the \gls{uav} is defined as a multiple of $\mathrm{E_{min}}$. \figurename~\ref{fig:uav_energy} denotes the impact of \gls{uav} energy budget on the trajectory optimization. As visible from the figure, when increasing the budget, the \gls{uav} is able to deviate further away from the shortest path trajectory. But eventually, it cannot go much further as it would risk not serving the users on the opposite side (as discussed previously). Once the \gls{uav} energy is sufficient to draw the optimal trajectory across the area, any further increase of the \gls{uav} energy would likely allow the \gls{uav} to slow down its speed or hover on the optimal location for a longer time, thus improving the transmission energy efficiency of the system. The minimum amount of energy required to reach the optimal trajectory is hence important to dimension the \gls{uav} battery capacity.
\subsection{Impact of \gls{ris} Position on \gls{uav} Transmission Power}
\begin{figure}[t]
    \centering
    \includegraphics[width=0.45\textwidth]{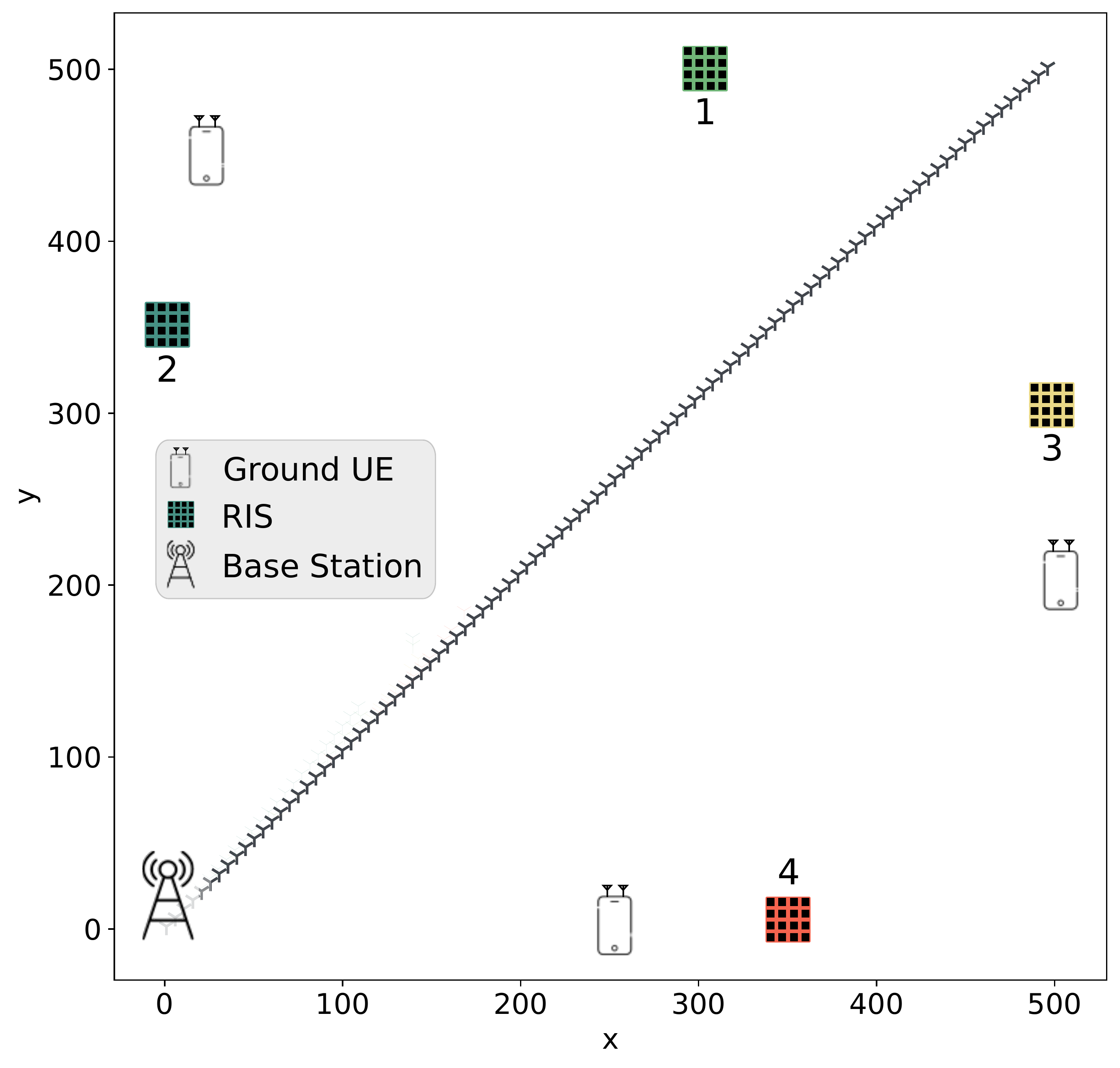}
    \caption{Different \gls{ris} positions configuration}
    \label{fig:ris_config}
\end{figure}
\begin{figure}[t!]
    \centering
    \includegraphics[width=0.45\textwidth]{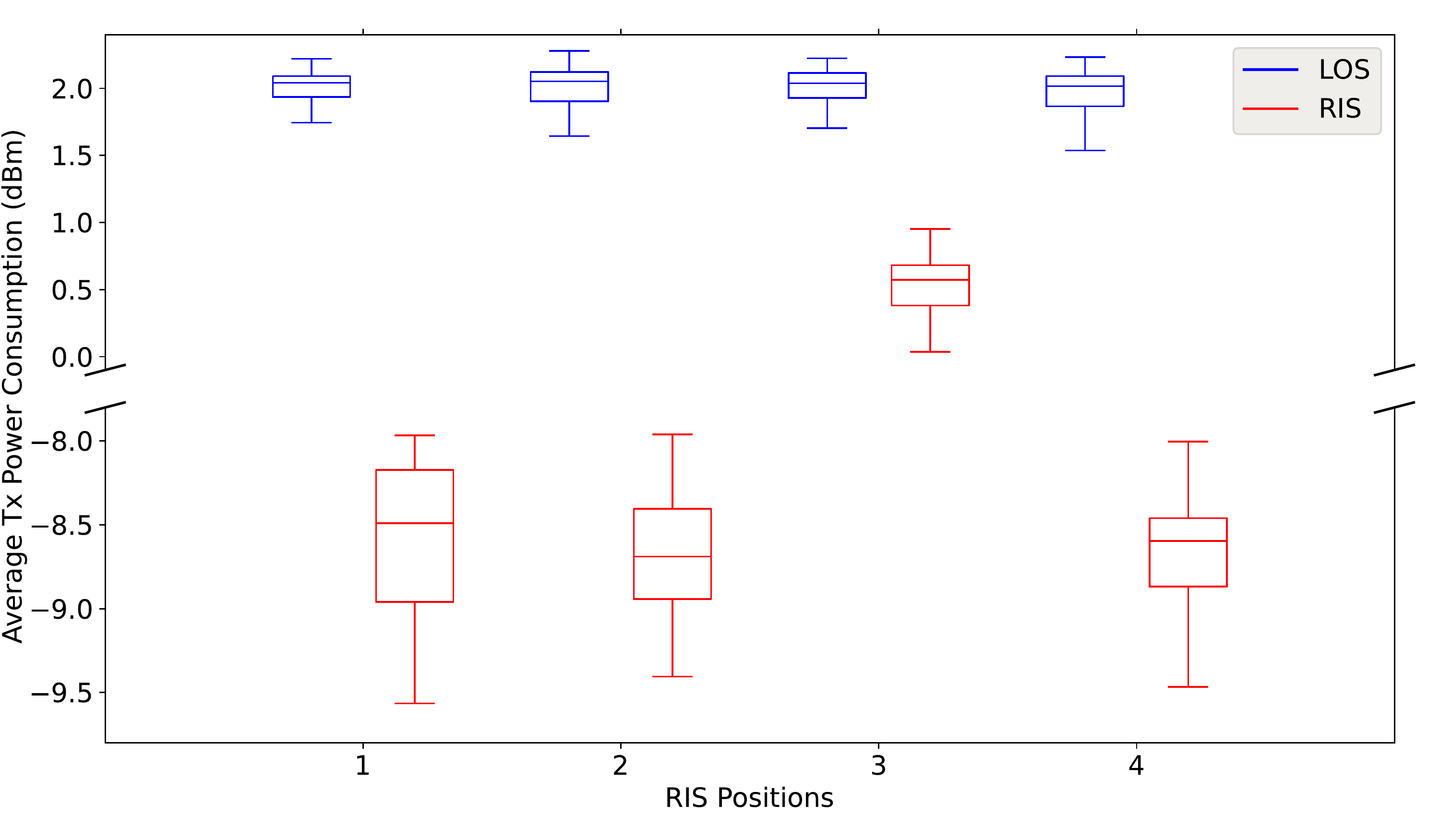}
    \caption{Average transmission power consumption of the \gls{uav} over the \gls{los} and \gls{ris} link for different positions of the \gls{ris} as shown in \figurename~\ref{fig:ris_config}, for 50 different network service requirements.}
    \label{fig:link_power}
\end{figure}
As concluded in previous subsection, the \gls{uav} has a tendency to move towards the \gls{ris}. To determine the impact of the \gls{ris} position on the \gls{uav} transmission power consumption, we obtained the optimal trajectory for different network service requirements i.e. spectral efficiency or $R_{min}$ per \gls{ue} randomly chosen between $0.01$ and $0.757$ bits/s/Hz. \figurename~\ref{fig:link_power} shows the boxplot for the average \gls{uav} power consumption over \gls{los} and \gls{ris} links for different positions of the \gls{ris}, as denoted in \figurename~\ref{fig:ris_config},
obtained for 50 different sets of values of $R_{min}$ randomly chosen between $0.01$ and $0.757$ bits/s/Hz.

As visible from the figure, the \gls{ris} position is crucial with respect to not only the \glspl{ue} but also the initial trajectory of the \gls{uav}. When in position 3, the \gls{ris} is closer to both the \glspl{ue} and the \gls{uav} initial trajectory and hence the optimization problem uses the \gls{ris} link to serve the \glspl{ue}. On the other hand, in position 2, the \gls{ris} is closer to the \gls{ue} but is further away from the \gls{uav} initial trajectory and hence the \gls{ris} link is not much used by the \gls{uav}. In the other two positions, the \gls{ris} is extremely far away from the \glspl{ue} and hence is also not much used. 

So the optimal solution is able to use the \gls{ris} to serve the users when the \gls{ris} is closer to the \glspl{ue} as well as the \gls{uav} initial trajectory. This also signifies that, the \gls{sca} is extremely sensitive to network configurations especially with respect to \gls{uav} initial trajectory and \gls{ris} and \glspl{ue} positions. It also highlights a drawback in the usage of \gls{sca}. By using \gls{sca} to determine the optimal \gls{uav} trajectory, \gls{ris} phase and \gls{uav} transmission power, it is really dependent on the initial state (a.k.a trajectory) we decided to solve the optimization problem. Hence, the usage of data driven methods such as reinforcement learning techniques to jointly optimize the \gls{uav} trajectory, \gls{ris} phase and \gls{uav} transmission power can be extremely lucrative and pursued for further work, as it would provide a more generalized solution regardless of the specific network configuration.

\section{Conclusion}\label{sec:conclusion}
Beyond 5G and 6G Networks are expected to provide a certain service level while reducing the power consumption of the system. To this end, we discussed the usage of \glspl{uav} and \gls{ris} as a way to guarantee certain service requirements while trying to minimise the power consumption of the system.

In this work, we devised jointly, a method to roughly optimize \gls{uav} trajectory, \gls{ris} phase and \gls{uav} transmission power consumption to provide a certain guaranteed service rate to the \glspl{ue} on the ground. We showed the usage of convex approximation techniques can provide a feasible solution.

Moving forward, the usage of reinforcement learning seems very attractive especially due to the sensitive nature of convex approximation schemes to different network configurations. 
\section*{Appendix I}
\label{sec:appendix}
In Appendix I, the channel models incorporated in the optimization problem are presented.
\subsubsection{\gls{bs} to \gls{uav} ($h^{BU}$) Channel} 
We assume a \gls{los} channel based on the UAV-UE channel model from~\cite{cai2020resource}. From \figurename~\ref{fig:problem_scanrio}, we devise $d^{\mathrm{BU}}[n] =\sqrt{\Vert \mathbf{Z}^{BS}-\mathbf{Z}[n]\Vert^{2}}$ as the Euclidean distances between \gls{uav} and \gls{bs} where $\mathbf{Z}[n] = [x_U[n],y_U[n],H_U]$ and $\mathbf{Z}^{BS} = [x_B,y_B,H_B]$ are the coordinates of \gls{uav} and \gls{bs} at a particular time instant $n$. So, the \gls{los} channel from \gls{bs} to \gls{uav} is designed as follows
\begin{align*} 
    \mathrm{h}^{\mathrm{BU}}[n]&=\sqrt{\frac{\alpha_0}{(d^{\mathrm{BU}}[n])^2}}[1,e^{-j\frac{2\pi\Delta_{\mathrm{B}x}}{\lambda_{\mathrm{c}}}\sin\theta^{\mathrm{BU}}\cos\xi^{\mathrm{BU}}},\\ &\ldots, e^{-j\frac{2\pi\Delta_{\mathrm{B}x}}{\lambda_{\mathrm{c}}}(M_{\mathrm{B}x}-1)\sin\theta^{\mathrm{BU}}_k\cos\xi^{\mathrm{BU}}}]^{\mathrm{H}}\\ &\otimes[1,e^{-j\frac{2\pi\Delta_{\mathrm{B}y}}{\lambda_{\mathrm{c}}}\sin\theta^{\mathrm{BU}}\sin\xi^{\mathrm{BU}}},\\ 
    &\ldots, e^{-j\frac{2\pi\Delta_{\mathrm{B}y}}{\lambda_{\mathrm{c}}}(M_{\mathrm{B}y}-1)\sin\theta^{\mathrm{BU}}\sin\xi^{\mathrm{BU}}}]^{\mathrm{H}}\in \mathbb{C}^{M_{\mathrm{B}}\times 1}\\
    \tag{36}
\end{align*}
where, $\mathbf{h}^{\mathrm{BU}}[n]$ is the channel vector based on the \gls{aod}.
$\Delta_{\mathrm{Bx}}$ and $\Delta_{\mathrm{By}}$ are the separation between antenna elements in x-direction and y-direction for \gls{ue}. Also, $M_{\mathrm{B}x}$ and $M_{\mathrm{B}y}$ is the number of antenna elements in x and y-direction for \gls{bs}, $M_{\mathrm{B}} = M_{\mathrm{B}x}\times M_{\mathrm{B}y}$ is total number of antenna elements for \gls{bs}  and $\lambda_\mathrm{c}$ is the carrier wavelength. $\theta^{\mathrm{BU}}$ and $\xi^{\mathrm{BU}}$ are the \gls{aod} for the link from \gls{bs} to the \gls{uav}. From \figurename~\ref{fig:angles}, it can be observed that $\sin\theta^{\mathrm{BU}} = \dfrac{\Vert H_B \Vert}{d^{\mathrm{BU}}}$, $\sin\xi^{\mathrm{BU}} = \dfrac{\Vert x_{B} - x_U[n] \Vert}{\Vert l_{B} - l_U[n]\Vert}$ and $\cos\xi^{\mathrm{BU}} = \dfrac{\Vert y_{B} - y_U[n] \Vert}{\Vert l_{B} - l_U[n]\Vert}$ where, $l_{B} = [x_{B},y_{B}]$ and $l_U[n] = [x_U[n],y_U[n]]$.

\subsubsection{\gls{uav} to \gls{ue} ($h^{UG}$) Channel}
For the link between \gls{uav} and \gls{ue}, which is \gls{los} and between \gls{uav} and \gls{ue} through \gls{ris}, we adopt the channel model from~\cite{cai2020resource}. The \gls{uav}, \gls{ris} and \gls{ue}, as previously mentioned, have a \gls{upa} antenna with $M_{\mathrm{U}},M_{\mathrm{R}}$ and $M_{\mathrm{G}}$ elements, respectively. Due to our XL-MIMO RIS assumption, the channel model we propose corresponds to each subsection/group of elements in the XL-MIMO RIS that we are using to serve different \glspl{ue}, similar to the approach proposed in~\cite{lin2020reconfigurable} to reflect sharp beams towards specific destinations. We assume that these groups have sufficient spatial separation thereby neglecting interference among them. From \figurename~\ref{fig:problem_scanrio}, we devise $d^{\mathrm{UR}} = \sqrt{\Vert \mathbf{Z}^{RIS}-\mathbf{Z}[n]\Vert^{2}},~d_{k}^{\mathrm{UG}} =\sqrt{\Vert \mathbf{Z}_{k}^{UE}-\mathbf{Z}[n]\Vert^{2}},~\mathrm{and}~d_{k}^{\mathrm{RG}} =\sqrt{\Vert\mathbf{Z}^{RIS}-\mathbf{Z}_{k}^{UE}\Vert^{2}}$ as the Euclidean distances between \gls{uav} and \gls{ris}, \gls{uav} and $k^{th}$\gls{ue} and \gls{ris} and $k^{th}$ \gls{ue} respectively, where $\mathbf{Z}^{RIS} = [x_R,y_R,H_R]$ and $\mathbf{Z}_k^{UE} = [x_{G,k},y_{G,k},0]$ are the coordinates of \gls{ris} and $k^{th}$ \gls{ue}. So, the \gls{los} channel from \gls{uav} to \gls{ue} is designed as follows,
\begin{align*} 
    \mathrm{h}^{\mathrm{UG}}_k[n]&=\sqrt{\frac{\alpha_0}{(d^{\mathrm{UG}}_k[n])^2}}[1,e^{-j\frac{2\pi\Delta_{\mathrm{U}x}}{\lambda_{\mathrm{c}}}\sin\theta^{\mathrm{UG}}_k\cos\xi^{\mathrm{UG}}_k},\\ &\ldots, e^{-j\frac{2\pi\Delta_{\mathrm{U}x}}{\lambda_{\mathrm{c}}}(M_{\mathrm{U}x}-1)\sin\theta^{\mathrm{UG}}_k\cos\xi^{\mathrm{UG}}_k}]^{\mathrm{H}}\\ &\otimes[1,e^{-j\frac{2\pi\Delta_{\mathrm{U}y}}{\lambda_{\mathrm{c}}}\sin\theta^{\mathrm{UG}}_k\sin\xi^{\mathrm{UG}}_k},\\ 
    &\ldots, e^{-j\frac{2\pi\Delta_{\mathrm{U}y}}{\lambda_{\mathrm{c}}}(M_{\mathrm{U}y}-1)\sin\theta^{\mathrm{UG}}_k\sin\xi^{\mathrm{UG}}_k}]^{\mathrm{H}}\in \mathbb{C}^{M_{\mathrm{U}}\times 1}\\
    \tag{37}
\end{align*}
where, $\mathrm{h}_k^{\mathrm{UG}}[n]$ is the channel vector based on the \gls{aod} and $\lambda_c$ is the carrier wavelength. $\theta^{\mathrm{UG}}_k$ and $\xi^{\mathrm{UG}}_k$ are the \gls{aod} for the link from \gls{uav} to $k^{th}$ \gls{ue}. From \figurename~\ref{fig:angles}, it can be observed that $\sin\theta^{\mathrm{UG}}_k = \dfrac{\Vert H_U \Vert}{d^{\mathrm{UG}}_k}$, $\sin\xi^{\mathrm{UG}}_k = \dfrac{\Vert x_{G,k} - x_U[n] \Vert}{\Vert l_{G,k} - l_U[n]\Vert}$ and $\cos\xi^{\mathrm{UG}}_k = \dfrac{\Vert y_{G,k} - y_U[n] \Vert}{\Vert l_{G,k} - l_U[n]\Vert}$ where, $l_{G,k} = [x_{G,k},y_{G,k}]$ and $l_U[n] = [x_U[n],y_U[n]]$.
\subsubsection{\gls{uav} to \gls{ris} to \gls{ue} ($h^{URG}$) Channel}
Similarly, the channel from \gls{uav} to \gls{ris} is defined as follows
\begin{align*} 
    \mathrm{H}^{\mathrm{UR}}[n]&=\mathrm{h}^{\mathrm{RU}}[n]\otimes(\mathrm{h}^{\mathrm{UR}}[n])^{\mathrm{H}}\\ &=\sqrt{\frac{\alpha_0}{(d^{\mathrm{UR}}[n])^2}}[1,e^{-j\frac{2\pi\Delta_{\mathrm{R}x}}{\lambda_{\mathrm{c}}}\sin\theta^{\mathrm{RU}}\cos\xi^{\mathrm{RU}}},\\ &\ldots, e^{-j\frac{2\pi\Delta_{\mathrm{R}x}}{\lambda_{\mathrm{c}}}(M_{\mathrm{R}x}-1)\sin\theta^{\mathrm{RU}}\cos\xi^{\mathrm{RU}}}]^{\mathrm{H}}\\ &\otimes[1,e^{-j\frac{2\pi\Delta_{\mathrm{R}y}}{\lambda_{\mathrm{c}}}\sin\theta^{\mathrm{RU}}\sin\xi^{\mathrm{RU}}},\\ 
    &\ldots, e^{-j\frac{2\pi\Delta_{\mathrm{R}y}}{\lambda_{\mathrm{c}}}(M_{\mathrm{R}y}-1)\sin\theta^{\mathrm{RU}}\sin\xi^{\mathrm{RU}}}]^{\mathrm{H}}\\ &\otimes[1,e^{-j\frac{2\pi\Delta_{\mathrm{U}x}}{\lambda_{\mathrm{c}}}\sin \theta^{\mathrm{UR}}\cos\xi^{\mathrm{UR}}},\\ &\ldots,e^{-j\frac{2\pi\Delta_{\mathrm{U}x}}{\lambda_{\mathrm{c}}}(M_{\mathrm{U}x}-1)\sin\theta^{\mathrm{UR}}\cos\xi^{\mathrm{UR}}}]\\ &\otimes[1,e^{-j\frac{2\pi\Delta_{\mathrm{U}y}}{\lambda_{\mathrm{c}}}\sin\theta^{\mathrm{UR}}\sin\xi^{\mathrm{UR}}},\\ &\ldots,e^{-j\frac{2\pi\Delta_{\mathrm{U}y}}{\lambda_{\mathrm{c}}}(M_{\mathrm{U}y}-1)\sin\theta^{\mathrm{UR}}\sin\xi^{\mathrm{UR}}}],\in \mathbb{C}^{M_{\mathrm{R}}\times M_{\mathrm{U}}}
    \tag{38}
\end{align*}
where, $\mathrm{h}^{\mathrm{RU}}[n]$ and $\mathrm{h}^{\mathrm{UR}}[n]$ are the channel vectors based on the \gls{aoa} and \gls{aod} respectively, $d^{\mathrm{UR}}[n]$ is the distance between the \gls{uav} and \gls{ris}, % add a definition for distance once the image is completed
$\Delta_{\mathrm{Rx}}$ and $\Delta_{\mathrm{Ry}}$ is the separation between antenna elements in x-direction and y-direction for \gls{ris} and $\Delta_{\mathrm{Ux}}$ and $\Delta_{\mathrm{Uy}}$ is the separation between antenna elements in x-direction and y-direction for \gls{uav}. Also, $M_{\mathrm{R}x}$ and $M_{\mathrm{R}y}$ is the number of antenna elements in x and y-direction for \gls{ris} and $M_{\mathrm{U}x}$ and $M_{\mathrm{U}y}$ are the number of antenna elements in x and y-direction for \gls{uav}, $\lambda_c$ is the carrier wavelength. $\theta^{\mathrm{RU}}$ and $\xi^{\mathrm{RU}}$ are the \gls{aoa} and $\theta^{\mathrm{UR}}$ and $\xi^{\mathrm{UR}}$ are the \gls{aod} for the link from \gls{uav} to \gls{ris}. From \figurename~\ref{fig:angles}, it can be observed that $\theta^{\mathrm{RU}} = \theta^{\mathrm{UR}}$ and $\xi^{\mathrm{RU}} = \xi^{\mathrm{UR}}$. So $\sin\theta^{\mathrm{RU}} = \sin\theta^{\mathrm{UR}} = \dfrac{\Vert H_U - H_R \Vert}{d^{\mathrm{UR}}}$, $\sin\xi^{\mathrm{RU}} = \sin\xi^{\mathrm{UR}} = \dfrac{\Vert x_R - x_U[n] \Vert}{\Vert l_R - l_U[n]\Vert}$ and $\cos\xi^{\mathrm{RU}} = \cos\xi^{\mathrm{UR}} = \dfrac{\Vert y_R - y_U[n] \Vert}{\Vert l_R - l_U[n]\Vert}$ where, $l_R = [x_R,y_R]$.
Subsequently, the channel from \gls{ris} to $k^{th}$ \gls{ue} is defined as follows
\begin{align*} 
    \mathrm{h}^{\mathrm{RG}}_k&=\sqrt{\frac{\alpha_0}{(d^{\mathrm{RG}}_k)^2}}[1,e^{-j\frac{2\pi\Delta_{\mathrm{R}x}}{\lambda_{\mathrm{c}}}\sin \theta^{\mathrm{RG}}_k\cos\xi^{\mathrm{RG}}_k},\\ &\ldots,e^{-j\frac{2\pi\Delta_{\mathrm{R}x}}{\lambda_{\mathrm{c}}}(M_{\mathrm{R}x}-1)\sin\theta^{\mathrm{RG}}_k\cos\xi^{\mathrm{RG}}_j}]^\mathrm{H}\\ &\otimes[1,e^{-j\frac{2\pi\Delta_{\mathrm{R}y}}{\lambda_{\mathrm{c}}}\sin\theta^{\mathrm{RG}}_k\sin\xi^{\mathrm{RG}}_k},\\ &\ldots,e^{-j\frac{2\pi\Delta_{\mathrm{R}y}}{\lambda_{\mathrm{c}}}(M_{\mathrm{R}y}-1)\sin\theta^{\mathrm{RG}}_k\sin\xi^{\mathrm{RG}}_k}]^\mathrm{H},\in \mathbb{C}^{M_{\mathrm{R}}\times 1}
    \tag{39}
\end{align*}
where, $\mathrm{h}^{\mathrm{RG}}[n]$ is the channel vector based on the \gls{aod} respectively, $d_k^{\mathrm{RG}}[n]$ is the distance between the \gls{ris} and \gls{ue},
$\Delta_{\mathrm{Rx}}$ and $\Delta_{\mathrm{Ry}}$is the separation between antenna elements in x-direction and y-direction for \gls{ris}. Also, $M_{\mathrm{R}x}$ and $M_{\mathrm{R}y}$ is the number of antenna elements in x and y-direction for \gls{ris} and $\lambda_c$ is the carrier wavelength. Additionally, $\theta^{\mathrm{RG}}$ and $\xi^{\mathrm{RG}}$ are the \gls{aod} for the link from \gls{ris} to $k^{th}$ \gls{ue}. From \figurename~\ref{fig:angles}, it can be observed that $\sin\theta^{\mathrm{RG}}_k = \dfrac{\Vert H_R \Vert}{d^{\mathrm{RG}}_k}$, $\sin\xi^{\mathrm{RG}}_k = \dfrac{\Vert x_{G,k} - x_R \Vert}{\Vert l_{G,k} - l_R\Vert}$ and $\cos\xi^{\mathrm{RG}}_k = \dfrac{\Vert y_{G,k} - y_R \Vert}{\Vert l_{G,k} - l_R\Vert}$. Additionally, the phase shift introduced in the reflected signal by \gls{ris} is defined as
\begin{align*} 
    \mathbf{\Phi}_{k}[n] = \mathrm{diag} (e^{j\mathbf{\Phi}_{1,1,k}[n]},\ldots, e^{j\mathbf{\Phi}_{m_{\mathrm{R}x}, m_{\mathrm{R}y}, k}[n]}, &\\ \ldots, e^{j\mathbf{\Phi}_{M_{\mathrm{R}x}, M_{\mathrm{R}y}, k}[n]})\in \mathbb{C}^{M_{\mathrm{R}}\times M_{\mathrm{R}}},
    \tag{40}
\end{align*}
where $\mathbf{\Phi}_{m_{\mathrm{R}x}, m_{\mathrm{R}y}, k}[n] \in [0,2\pi), m_{\mathrm{R}x} = \{1,\ldots,M_{\mathrm{R}x}\},m_{\mathrm{R}y} = \{1,\ldots,M_{\mathrm{R}y}\}$ represents the phase control introduced to the $({m_{\mathrm{R}x}, m_{\mathrm{R}y}})^{th}$ reflecting  element  of  the  \gls{ris}. Hence, end-to-end effective channel between the \gls{uav} and the $k^{th}$ \gls{ue} reflected by the \gls{ris} is given by
\begin{equation*} 
    (\mathrm{H}^{\mathrm{URG}}_{k}[n])^{\mathrm{H}}=(\mathrm{h}_{k}^{\mathrm{RG}}[n])^{\mathrm{H}}\mathbf{\Phi}_{k}[n]\mathrm{H}^{\mathrm{UR}}[n]\in \mathbb{C}^{1\times M_{\mathrm{U}}}.
    \tag{41}
\end{equation*}
\bibliographystyle{IEEEtran}
\bibliography{bibliography2.bib}

\end{document}